\documentclass{aastex6}
\usepackage{multirow}
\usepackage{color}

\fullcollaborationName{The Friends of AASTeX Collaboration}
\begin{document}




\title{A new look at the molecular gas in M42 and M43; possible evidence for cloud-cloud collision which triggered formation of the OB stars in the Orion Nebula Cluster}

\author{Yasuo Fukui\altaffilmark{1}, Kazufumi Torii\altaffilmark{2}, Yusuke Hattori\altaffilmark{1}, Atsushi Nishimura\altaffilmark{1}, Akio Ohama\altaffilmark{1}, Yoshito Shimajiri\altaffilmark{3}, Kazuhiro Shima\altaffilmark{4}, Asao Habe\altaffilmark{4}, Hidetoshi Sano\altaffilmark{1}, Mikito Kohno\altaffilmark{1}, Hiroaki Yamamoto\altaffilmark{1}, Kengo Tachihara\altaffilmark{1}, and Toshikazu Onishi\altaffilmark{5}}

\affil{$^1$Department of Physics, Nagoya University, Chikusa-ku, Nagoya, Aichi 464-8601, Japan}
\affil{$^2$Nobeyama Radio Observatory, 462-2 Nobeyama Minamimaki-mura, Minamisaku-gun, Nagano 384-1305, Japan}
\affil{$^3$Laboratoire AIM, CEA/DSM-CNRS-Universit$\acute{\rm e}$ Paris Diderot, IRFU/Service d’Astrophysique, CEA Saclay, F-91191 Gif-sur-Yvette, France}
\affil{$^4$Faculty of Science, Department of Physics, Hokkaido University, Kita 10 Nishi 8 Kita-ku, Sapporo 060-0810, Japan}
\affil{$^5$Department of Astrophysics, Graduate School of Science, Osaka Prefecture University, 1-1 Gakuen-cho, Nakaku, Sakai, Osaka 599-8531, Japan}


\begin{abstract}
The Orion Nebula Cluster toward the H{\sc ii} region M42 is the most outstanding young cluster at the smallest distance 410\,pc among the rich high-mass stellar clusters. By newly analyzing the archival molecular data of the $^{12}$CO($J$\,=\,1--0) emission at 21$''$ resolution, we identified at least three pairs of complementary distributions between two velocity components at 8\,km\,s$^{-1}$ and 13\,km\,s$^{-1}$. We present a hypothesis that the two clouds collided with each other and triggered formation of the high-mass stars, mainly toward two regions including the nearly ten O stars, $\theta^{1}$\,Ori and $\theta^{2}$\,Ori, in M42 and the B star, NU\,Ori, in M43. The timescale of the collision is estimated to be $\sim$\,0.1\,Myr by a ratio of the cloud size and velocity corrected for projection, which is consistent with the age of the youngest cluster members less than 0.1\,Myr. The majority of the low-mass cluster members were formed prior to the collision in the last one Myr. We discuss implications of the present hypothesis and the scenario of high-mass star formation by comparing with the other eight cases of triggered O star formation via cloud-cloud collision.
\end{abstract}





\section{Introduction} \label{sec:intro}

High-mass stars are influential in the evolution of galaxies by injecting a large amount of energy of ultraviolet photons, stellar winds and supernova explosions. It is of fundamental importance to understand the formation mechanism of high-mass stars in order to elucidate structure and evolution of galaxies, and considerable efforts have been devoted to understand the formation mechanism of high-mass stars. It has been discussed that self-gravitational massive aggregation of molecular gas or protostars are probable sites of high-mass star formation. The theories of high-mass star formation include the monolithic collapse and the competitive accretion \citep[for recent reviews see][]{zin2007, tan2014}, and massive and dense clouds including the hot cores and the infrared dark clouds are studied as possible sites of such high-mass star formation \citep[e.g.,][]{ega1998}. It is however puzzling that these massive objects are not commonly associated with H{\sc ii} regions, the most obvious signature of O\,/\,early B star formation, and we do not yet have convincing confrontation between the theories and observations \citep{tan2014}.

Recently, supersonic collision between the interstellar molecular clouds is discussed as a possible triggering mechanism of high-mass star formation both observationally and theoretically. Up to now we have 8 regions of cloud-cloud collision toward the super star clusters and the H{\sc ii} regions in the Milky Way disk, and toward the young O stars in the LMC \citep{fur2009, oha2010, tor2011, tor2015, fuk2014, fuk2015b, fuk2016a, sai2017}. These collisions take place supersonically at a velocity of 10\,--\,30\,km\,s$^{-1}$, and the high velocity indicates that the cloud motion is not governed by the cloud-cluster self-gravity. 
{
Subsequently, we find similar cases of cloud-cloud collision triggering O/early B star formation is still increasing in number  \citep{fur2009, oha2010, fuk2014, fuk2016a, 2018arXiv170606002N, 2018arXiv170606956N, 2018arXiv170605771F, 2018arXiv170605768F, nak2012, 2017arXiv170605652O, 2017arXiv170605664T, 2017arXiv170605871H, 2017arXiv170605763S, 2014ApJ...797...58D, tor2011, tor2017, 2017arXiv170902320O, tor2015, 2017arXiv170605659O, 2017arXiv170607164T, 2016ApJ...819...66D, 2017ApJ...834...22D, 2013ApJ...768...72S, 2017arXiv170607964K, 1986ApJ...305..353M, 2009PASJ...61...39M, 2001PASJ...53..793O, fujita2017, 2017ApJ...835L..14G, 2017arXiv170808149S, 2017ApJ...837...44D, 2017ApJ...840..111T, 2016ApJ...833...85B, 2011ApJ...731...23S, 1994ApJ...429L..77H, 2000ApJ...535..857S, 2015PASJ...67..109T, 2017PASJ...69L...5F, fuk2015b, sai2017}.
}
It is shown by theoretical works that cloud-cloud collision increases the cloud density and the effective sound speed in the shocked interface layer \citep{hab1992, whi2004, ana2010, tak2014}. According to the recent magnetohydrodynamical numerical simulations for a head-on collision at 20\,km\,s$^{-1}$ by \citet{ino2013}, the mass accretion rate is enhanced by two orders of magnitude to $10^{-4}$\,--\,$10^{-3}$\,$M_\odot$\,yr$^{-1}$ due to the collision as compared with the pre-collision state. The mass accretion rate $10^{-4}$\,--\,$10^{-3}$\,$M_\odot$\,yr$^{-1}$ satisfies the theoretical requirement to overcome the radiation pressure feedback of the forming O star \citep{wol1986, mck2003}. The typical collision time scale is estimated to be 0.1\,Myr from a ratio of the O star cluster size and velocity, 1\,pc\,/\,10\,km\,s$^{-1}$, and stellar mass of 10\,--\,100\,$M_\odot$ can be accumulated at the enhanced mass accretion rate in the timescale.

Currently, a key question is if cloud-cloud collision is one of the major mechanisms of high-mass star formation. It is premature to conclude that cloud-cloud collision is common because the number of samples is yet too small. One of the difficulties in studying high-mass star formation is that the number of high-mass star forming regions is small as compared with low-mass star formation sites. The nearest young O star cluster is the Orion Nebula Cluster (ONC) including M42 and M43 located at 410\,pc from the sun \citep[see for reviews][]{mue2008, ode2008, pet2008}. The ONC has been best studied among the O star clusters at various wavelengths from X rays to radio emissions and is the primary object where theories of high-mass star formation are tested. The other known regions of high-mass star formation are located at a distance around 2\,kpc or larger.

The Orion A molecular cloud which harbors the ONC is an active star forming region, and includes OMC(Orion Molecular Cloud)-1, OMC-2, and OMC-3 in the M42\,--\,M43 region.  A number of molecular line observations were made toward the Orion A cloud in the last few decades mainly in the millimeter/sub-millimeter CO emission lines \citep{kut1977, mad1986, sug1986, tak1986, bal1987, cas1990, hey1992, tat93, sak1994, whi1995, nag1998, plu2000, shi2011, nak2012, wil2005, rip2013, shi14, nis2015}. $\sim$\,100 OB stars and ~2000 low mass young stars are distributed in and around the molecular cloud having $10^{5}$\,$M_\odot$. In particular, the H{\sc ii} region M42 is the most active site of O star formation which includes about ten O\,/\,early B stars. On a large scale including the Orion B cloud, four OB associations named Ori OB1a, Ori OB1b, Ori OB1c, and Ori OB1d are identified \citep{bla1991} and they may be distributed spatially in age sequence. A model of the sequential formation of the OB associations was therefore proposed, where OB stars are formed due to triggering by the ionization-shock fronts \citep{elm1977}. The Orion A cloud appears to be converging toward M42 showing a V shape pointing to the north and a model consisting of self-gravitationally-bound converging motion was presented to explain the shape and formation of the ONC \citep{har2007}, whereas details of the star formation remain elusive \citep[see also for low-mass star formation,][]{stu2016}. 

In this paper we present a new analysis of the archival data of $^{12}$CO($J$\,=\,1--0) transition obtained with large single-dish telescopes aiming to test the scenario of cloud-cloud collision as a possible mechanism of high-mass star formation. The present paper focuses on the distribution and kinematics of the cloud. In Section 2 we give a summary of observational and theoretical papers on cloud-cloud collision and present common signatures characteristic to the collision scenario. Section 3 explains an analysis of the CO data, and Section 4 presents a model of the colliding clouds along with discussion. Section 5 gives discussion from a broader perspective and Section 6 concludes the paper.

\section{Cloud-cloud collision; observational and theoretical signatures}
Evidence for triggered formation of O\,/\,early B star(s) by cloud-cloud collision is found in the four super star clusters, Westerlund\,2, NGC\,3603, RCW\,38 and [DBS2003]\,179 \citep[][Kuwahara et al. 2017, in preparation]{fur2009, oha2010, fuk2014, fuk2016a}, in the two H{\sc ii} regions with single O stars, RCW\,120 and M20 (\citealt{tor2011, tor2015}, 2016), and in the single high-mass stars in the LMC, N159W-South and the Papillon-Nebula YSO of N159E \citep{fuk2015b, sai2017}. The physical parameters of the 8 objects along with the ONC are summarized in Table\,\ref{table:1}. The four super star clusters include 10\,--\,20 O stars with age of 0.1\,--\,4.0\,Myrs, while the others with a single O star have age less than 1.0\,Myr. Theoretical studies of cloud-cloud collision are made by using hydrodynamical numerical simulations \citep{hab1992, ana2010, ino2013, tak2014}. These studies allow us to gain an insight into the observational/theoretical signatures of colliding clouds. It is also notable that synthetic observations present observable characteristics of cloud-cloud collision including the complementary distribution at the two velocities and the bridging feature between the two colliding clouds \citep{haw2015b, haw2015, tor2017}. We shall summarize the major properties obtained by these previous studies in the following.

\subsection{The supersonic motion}
The observed cloud-cloud collision indicates the relative velocity between the two clouds is supersonic in a range of 10\,--\,30\,km\,s$^{-1}$, except for a few cases where the relative velocity is apparently small probably due to projection (N159W-South and N159E-Papillon). Such high velocity is gravitationally unbound by the typical cloud mass within a 10\,pc radius of the formed O star(s), $10^{3}$\,--\,$10^{4}$\,$M_\odot$. Observational verification of heating by O stars/H{\sc ii} regions is used to establish the physical association of the clouds with the formed O stars in spite of their large velocity difference \citep[e.g.,][]{fuk2014} which is otherwise interpreted as due to different kinematic distance. The number density of the molecular clouds in the Galactic spiral arms is high enough to cause frequent cloud-cloud collision every 8\,Myr for a giant molecular cloud as demonstrated by numerical simulations \citep{fuj2014, dob2015}.

The magnetohydrodynamical numerical simulations demonstrate that in the shock-compressed interface layer between the two clouds the density increases by a factor of ~10 and the turbulent velocity and Alfv\'en velocity increase by a factor of 5 \citep{ino2013}. This is because the shock front is deformed into disordered turbulence by density inhomogeneities of the initial clouds and the turbulence amplifies the field strength. The mass accretion rate is proportional to the third power of the effective sound speed including the turbulent velocity and Alfv\'en velocity, and the supersonic motion produces a high-mass accretion rate in the order of $10^{-4}$\,--\,$10^{-3}$\,$M_\odot$\,yr$^{-1}$, which is large enough to overcome the stellar radiation pressure and allow mass growth of the star(s). We note that the turbulent velocity field in the interface layer is nearly isotropic, providing a signature of the collision even when the velocity between the colliding clouds becomes small by the projection effect as shown in the interface layer of N159W-South and N159E-Papillon \citep{fuk2015b, sai2017}.

We see in Table\,\ref{table:1} that the number of forming O stars has a threshold value of molecular column density \citep{fuk2016a}. 10\,--\,20 O star formation takes place for high column density arround $10^{23}$\,cm$^{-2}$ derived from $^{12}$CO($J$\,=\,1--0) line intensity by using the X$_{CO}$ factor, whereas formation of a single O star happens for lower column density around $10^{22}$\,cm$^{-2}$. O star(s) is formed if one of the two clouds, not always both of them, satisfies these conditions. We suggest that in the high column density case the O star formation takes place in a small path length of the collision, e.g., $\sim$\,0.1\,pc \citep[see RCW\,38,][]{fuk2016a}, and that in the low column density case the path length prior to the O star formation is longer, $\sim$\,1\,pc \citep[see RCW\,120,][]{tor2015}

\subsection{The cloud dispersal by ionization and collisional interaction}
Molecular clouds within 1\,--\,10\,pc of O star formation will be rapidly dispersed by the ionization/winds and may not be observable at a timescale larger than $\sim$\,2\,Myr \citep{fuk2016a}, while the details will depend on the number of O stars, i.e., the ultraviolet radiation field, and the gas density distribution. Collisional interaction also converts the natal gas into the shocked interface layer and stars, and dissipates the natal clouds as discussed further in Section 2.3.

\begin{figure}
\epsscale{0.55}
\plotone{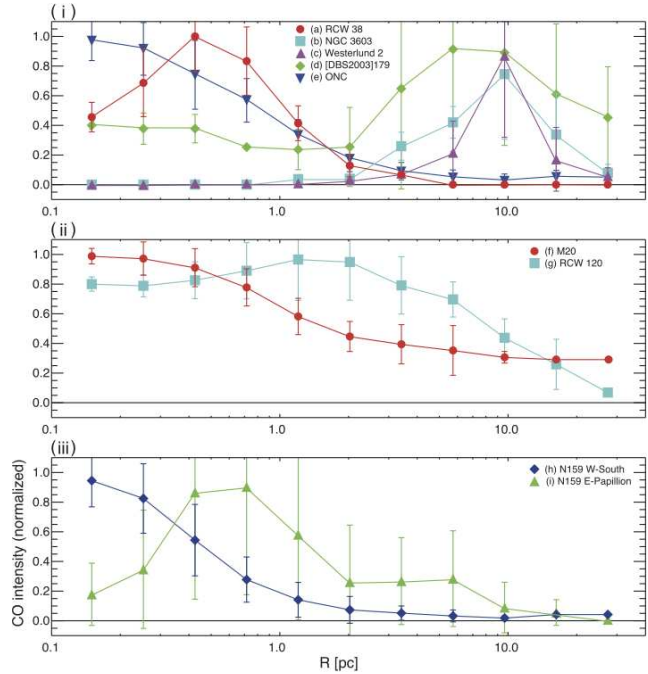}
\caption{CO radial distribution in 9 objects including the 8 regions of cloud-cloud collision and the Orion Nebula Cluster.
Radial distributions of the molecular emission in the nine regions from (a) to (i) listed in Table 1. (a) RCW\,38, (b) NGC\,3603, (c) Westerlund\,2, (d) [DBS2003]\,179, (e) the ONC, (f) RCW\,120, (g) M20, (h) N159W-South, and (i) N159E-Papillon Nebula YSO. 
The 5 regions in (i) have O stars over $\sim$\,10, and the remaining 4 regions in (ii) and (iii) have a single O star. 
The averaged intensity of molecular emission is calculated for circular areas at each radius from the center of the cluster or the O star. 
The CO transition and the velocity range used are listed in Table 1. 
The error bar corresponds to the $\pm$\,1\,$\sigma$ fluctuations in each grid.
   \label{fig1}}
\end{figure}

Figure\,\ref{fig1} shows the radial distribution of the associated molecular emission in the nine regions of O star formation in Table\,\ref{table:1}. {Table\,\ref{table:2} shows velocity ranges of the molecular emission in these regions.} The super star clusters including 10\,--\,20 O stars in a volume smaller than 1\,pc show often a central cavity of 10\,pc radius as seen in Westerlund\,2, NGC\,3603, and [DBS2003]\,179, whose ages are 2\,--\,4\,Myr. Similar cavities are observed in the other super star clusters including Arches, Quintuplet, and R\,136 \citep[e.g.,][]{sto2015, dem2011}. The only exception is RCW\,38 which shows rich molecular gas at a\,pc scale. A close look at RCW\,38 indicates a cavity of 0.5\,pc radius in the molecular gas created by the ionization, and the small ionization cavity likely reflects the small age 0.1\,Myr \citep{fuk2016a}. These super star clusters suggest that the ionization propagates at a typical velocity of $\sim$\,5\,km\,s$^{-1}$ (a ratio 0.5\,pc\,/\,0.1\,Myr, or 10\,pc\,/\,2\,Myr). This velocity is consistent with theoretical values of a ionization shock front \citep[e.g.,][]{elm1977}. We expect that the clouds within 1\,pc will soon be ionized and dispersed in another 0.1\,Myr, leading to termination of star formation in a few times 0.1\,Myr. The age spread of the cluster members formed by triggering is therefore short in the order of 0.1\,Myr in super star clusters. Such a small age spread is consistent with that of 0.1\,--\,0.3\,Myr of the young cluster members estimated by careful VLT and HST observations of the two super star clusters NGC\,3603 and Westerlund\,1 \citep{kud2012}.
	
Among the regions of single O star formation, the three regions M20, N159W-South and N159E-Papillon show rich molecular gas within 1\,pc and RCW\,120 shows a molecular cavity of 2\,pc radius. The size of the molecular cavity in single O star formation seems to be smaller than in the superstar clusters perhaps due to less ionizing photons. M20 whose age is 0.3\,Myr is extended over 3\,pc, whereas it also shows a clear cavity of neutral matter with a 0.1\,pc radius centered on the O star in the HST image (ESA/Hubble \& NASA, Bruno Conti). The Papillon-Nebula YSO in N159E shows a molecular cavity of 0.5\,pc radius, whose inside is filled by the H{\sc ii} region and the age of the Papillon-Nebula YSO is estimated to be 0.2\,Myr \citep{sai2017}. N159W-South has no H{\sc ii} region and no molecular cavity, and is driving protostellar outflow with the associated accretion disk \citep{{fuk2015b}}. N159W-South has a small age of 0.1\,Myr, and is younger than the Papillon-Nebula YSO. So, the molecular cavities in M20 and the Papillon-Nebula are likely created by the ionization. It is suggested that the molecular cavity in RCW\,120 was created by the cloud-cloud collision as first modeled by \citet{hab1992}, and the effect of the ionization is not dominant in forming the cavity (\citealt{tor2015}; see also for a different scenario \citealt{zav2010}). The single O star forming regions in the current sample are fairly young and the effect of the ionization does not seem to be prevailing, whereas the cloud dispersal by ionization will become dominant later.

\subsection{Complementary distribution between the colliding clouds}
It is often the case that colliding clouds are not of the same size as simulated by \citet{hab1992}, \citet{ana2010}, and \citet{tak2014}. If one of the colliding clouds is smaller than the other, the small cloud can create a cavity in the large cloud. This produces complementary distribution of the two clouds at different velocities, which is used to identify candidates for colliding clouds unless the cloud dispersal by the ionization is significant.

\begin{figure}
  \epsscale{1.}
   \plotone{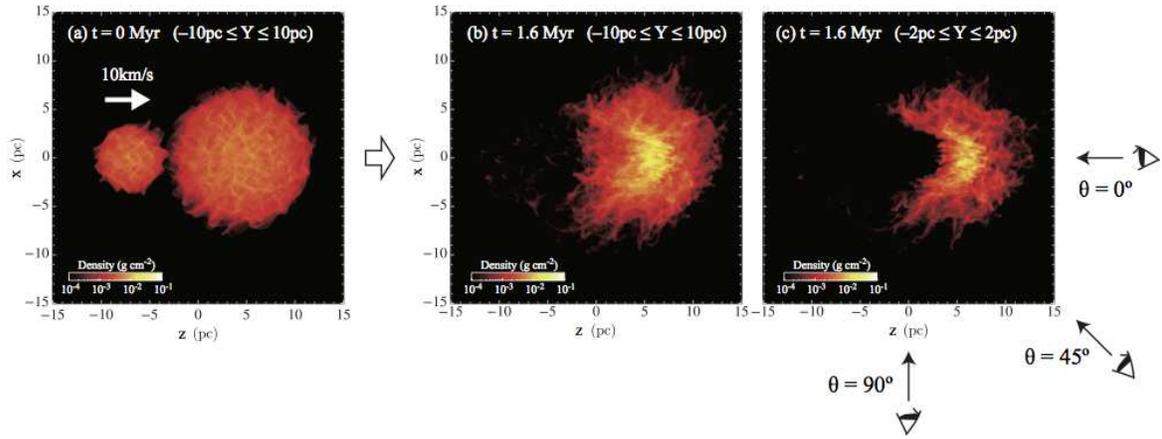}
  \caption{{Surface density plots of the 10\,km\,s$^{-1}$ collision model calculated by \citet{tak2014}. In (a), top-view of the two clouds prior to the collision is presented, while in (b) and (c) the snapshots at 1.6\,Myr after the onset of the collision are shown, where the integration ranges in the y-axis are $-15$--$+15$\,pc for (b) and $-1$--$+1$\,pc for (c). The eye symbols and arrows define the viewing angles used in the analyses of the synthetic $^{12}$CO($J$=1--0) data presented in Figures\,\ref{fig3}--\ref{fig5}.}
   \label{fig:takahira}}
\end{figure}

\begin{figure}
  \epsscale{.7}
   \plotone{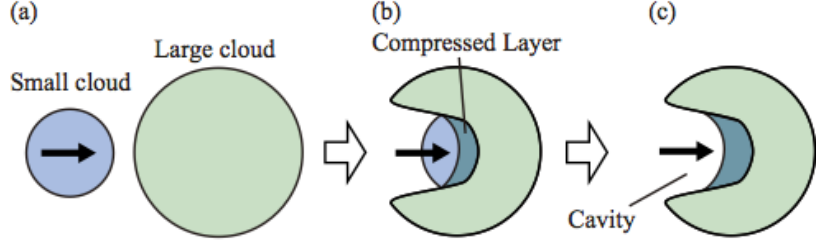}
  \caption{{Schematic picture of the cloud-cloud collision between two spherical clouds with different sizes, as simulated by \citet{tak2014} shown in Figure\,\ref{fig:takahira}. When the small cloud drives into the large cloud, a cavity is created in the large cloud, and the small cloud streams into the compressed layer formed at the interface of the collision. Compared with Figure\,\ref{fig:takahira}, the 0\,Myr and 1.6\,Myr cases correspond to (a) and (c) in this schematic, respectively.}
   \label{fig:takahira_sch}}
\end{figure}

In order to visualize the collision, we adopt the numerical simulations by \citet{tak2014}, and Table\,\ref{table:4} lists the simulation parameters. 
{In Figure\,\ref{fig:takahira} surface density plots of the 10\,km\,s$^{-1}$ collision model are presented, while in Figure\,\ref{fig:takahira_sch} a schematic picture is shown.}
In the model where the two clouds are spherical, the cross section of the cavity is determined by the size of the small cloud and the length of the cavity by the travel distance since the initiation of the collision.  The cavity is identified observationally as an intensity depression in the molecular distribution at the velocity of the large cloud. The projected distributions of the small cloud and the intensity depression are complementary, and they are in general displaced with each other because the angle of the relative motion to the line of sight, $\theta$, is not $0^\circ$. 

\begin{figure}
  \epsscale{1.}
   \plotone{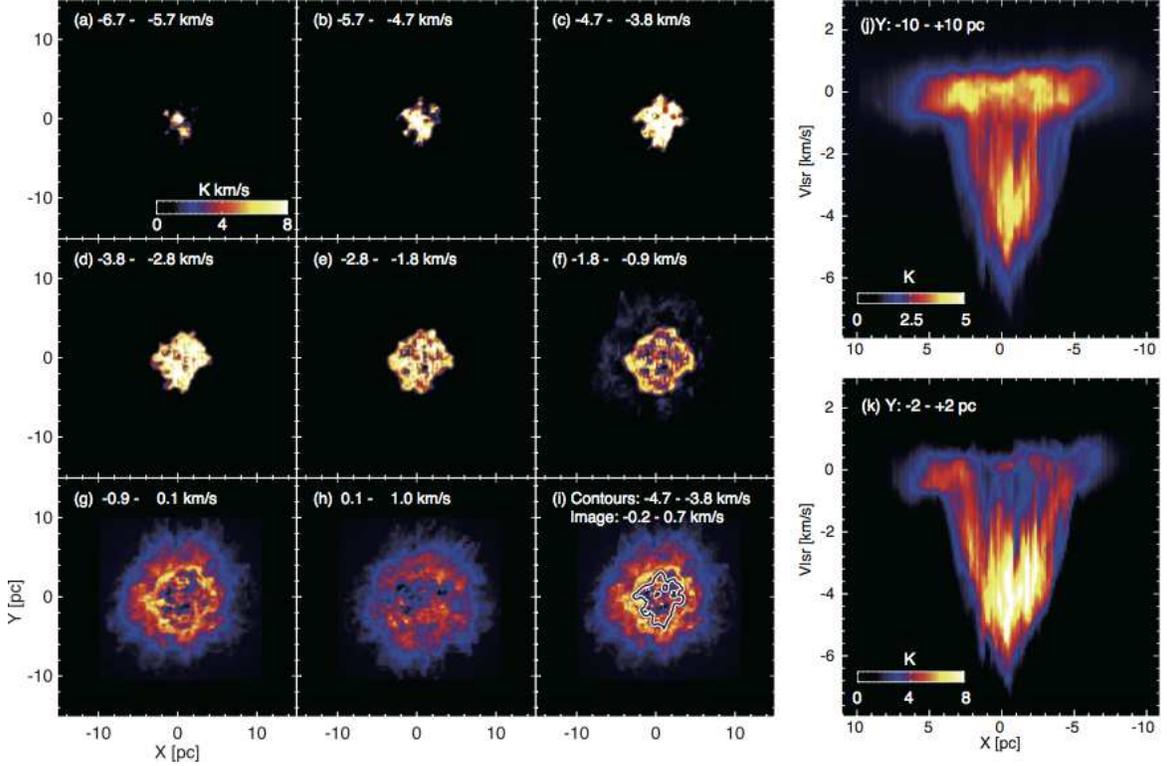}
  \caption{Synthetic observations of $^{12}$CO($J$=1--0) emission based on the numerical simulations by \citet{tak2014} observed at an angle of the relative motion to the line of sight $\theta=0^\circ$ (Figure\,\ref{fig:takahira}(c)).
  The parameters of the model are shown in Table\,\ref{table:4}. 
  (a)--(h) show the velocity channel distributions every 0.93\,km\,s$^{-1}$ in a velocity interval indicted in each panel. 
  (i) shows a complementary distribution between the large cloud, the image in (g), and the small cloud with the contour of (c) at 4\,K\,km\,s$^{-1}$.
  (j) and (k) show the position-velocity diagrams integrated over Y ranges of $-10$--$+10$\,pc and $-2$--$+2$\,pc, respectively.
   \label{fig3}}
\end{figure}

While the real cloud shape can be more complicated, it is still worthwhile to use the simple two-sphere model to gain an insight into the cloud distribution created by the collision. The relative motion of the colliding clouds generally has an inclination angle to the line of sight, and three cases of the inclination angle $\theta$, $0^\circ$, $45^\circ$, and $90^\circ$, are shown in the velocity channel distributions in Figures\,\ref{fig3}--\ref{fig5}. 

For $\theta=0^\circ$ (Figure\,\ref{fig3}) we see two velocity features whose distribution varies with velocity. We identify the initial two clouds at $-5.1$\,--\,$-3.1$\,km\,s$^{-1}$ (panels (b)--(d) in Figure\,\ref{fig3}) and $-1.2$\,--\,$0.7$\,km\,s$^{-1}$ (panels (f)--(h) in Figure\,\ref{fig3}) and the intermediate velocity feature mainly at $-3.1$\,--\,$-1.2$\,km\,s$^{-1}$ (panels (d)--(f) in Figure\,\ref{fig3}). The small cloud fits well with the cavity created in the large cloud by collision, showing the complementary distribution between them (Figure\ref{fig3}(i)). For $\theta=45^\circ$ (Figure\,\ref{fig4}) we see the two clouds at $-4.1$\,--\,$-2.2$\,km\,s$^{-1}$ (panels (c)--(d) in Figure\,\ref{fig4}) and $-1.2$\,--\,$0.7$\,km\,s$^{-1}$ (panels (f)--(g) in Figure\,\ref{fig4}) and the intermediate velocity feature mainly at $-2.2$\,--\,$-1.2$\,km\,s$^{-1}$ (panels (e) in Figure\,\ref{fig4}). In Figure\,\ref{fig4} we see a displacement of the complementary distribution between the cavity and the small cloud, which is a natural result of projection for $\theta=45^\circ$. 
The interaction between the two clouds mixes up the two clouds because of the momentum exchange and deforms the distribution of the two clouds in the intermediate velocity ranges; these are included in $-2.8$-- $-0.9$ km s$^{-1}$ (Figures \ref{fig3}e and \ref{fig3}f) and in $-2.2$-$-0.2$ km s$^{-1}$ (Figures \ref{fig4}e and \ref{fig4}f).
If the relative motion is vertical to the line of sight $\theta=90^\circ$ (Figure\,\ref{fig5}), we do not see the two velocity components. 
{For these three cases of $\theta$, the position-velocity diagrams with two integration ranges in the Y-axis of $\pm10$\,pc and $\pm$2\,pc are presented in Figures\,\ref{fig3}(j)--(k), \ref{fig4}(j)--(k), and \ref{fig5}(d)--(e). In the $0^\circ$ and $45^\circ$ cases, the two clouds are continuously distributed in velocity due to the $``$bridging$"$ feature (Section 2.4).}
Figures\,\ref{fig3}--\ref{fig5} illustrate that, even if we see only a singly-peaked velocity component in observations, a possibility of cloud-cloud collision is not excluded. 
{Figure \ref{figmom1} shows another presentation of the velocity distribution in first moment distributions for the three $\theta$s. Figures \ref{figmom1}a and \ref{figmom1}b indicate that the two clouds are clearly seen at two cloud velocities, suggesting that the first moment serves as another way to identify the two clouds, whereas the displacement at $\theta=45$ deg. is masked by averaging. 
}

In regions of single O star formation, the most significant complementary distribution toward an O star at a 0.1\,pc scale is found in M20 \citep{tor2017}. In RCW\,120 the dispersal of the small cloud is significant and only the large cloud with the intensity depression, one of the complementary clouds, is seen along with the remnant of the small cloud outside the ``bubble'' \citep{tor2015}. In N159 it is possible that the angular resolution of ALMA, 0.3\,pc\,$\times$\,0.2\,pc, is not high enough to resolve the complementary distribution which is likely to be in a scale of 0.1\,pc.  In the super star clusters where the ionization is more significant, it is probable that the complementary distribution is dispersed more quickly. Even in the youngest cluster RCW\,38 we see the ionization is dominant in shaping the clouds \citep{fuk2016a}. We note that the oldest cluster [DBS2003]\,179 among the four shows a large-scale complementary distribution at a scale more than 10\,pc, which is not affected much by the ionization (Kuwahara et al. 2017, in preparation). To summarize, the colliding clouds with small age and relatively low ionization are the most probable site to find the complementary distribution, whereas it is possible that the complementary distribution is often dispersed by the stellar feedback and the collisional interaction, depending on the cloud distribution. 

Another possible signature of cloud-cloud collision is seen in the optical/near-infrared extinction toward the H{\sc ii} region. In cloud-cloud collision the blue-shifted cloud is always on the far-side of the red-shifted cloud prior to the collision but moves to the near-side after the collision. As a result, one may expect that only the blue-shifted cloud, not the red-shifted cloud, is located in front of the H{\sc ii} region and can create dark heavily obscured features. Such dark lanes are in fact observed in M20 where the collision took place $\sim$\,0.3\,Myr ago to form a single O star; the dark lanes toward M20 show a good correspondence with the blue-shifted cloud \citep[][2016]{tor2011}. A caveat is that this may not always be the case if the collision happened very recently where the line-of-sight separation between the two clouds is still small. The super star cluster RCW\,38 shows a narrow dark lane in the near infrared as IRS\,1, whereas IRS\,1 is identified as part of the red-shifted cloud \citep{fuk2016a}. The blue-shifted cloud in RCW\,38 also corresponds to the near-infrared dark features surrounding the cluster. These authors interpret that IRS\,1 remains neutral without full ionization due to the small age of the system; the O star formation occurred only 0.1\,Myr ago and part of the red-shifted cloud within 1\,pc of the cluster center is still surviving in front of or inside the H{\sc ii} region. So, the dark lanes can arise from the both clouds depending on the age and shape of the clouds, and we need to be cautious in examining the extinction features when the collision is a very recent event.

\subsection{The bridging features between the two colliding clouds}
Numerical simulations show that the two colliding clouds become connected in velocity by a bridging feature in synthetic observations if the projected velocity separation is larger than the linewidths of the clouds \citep{haw2015b, haw2015}. The shocked interface layer appears as the bridging feature connecting the clouds in the position-velocity plots as shown in Figures\,\ref{fig3}(j)--(k) and \ref{fig4}(j)--(k). The bridging features are observable in the spots of collision, although all of them are not necessarily related to O star formation as shown later. 
As discussed in the previous subsection, the intensity depression in the large cloud due to creation of the cavity is also seen in the spots of collision, and if we make a position-velocity diagram so that a colliding spot is sliced, both of bridging features and intensity depression of the large cloud are seen, resulting in a ``V-shape'' gas distribution in the position-velocity diagram, as shown in Figures\,\ref{fig3}(k) and \ref{fig4}(k). In addition, in the $\theta = 45^\circ$ case in Figure\,\ref{fig4}(k), the V-shape becomes skew due to inclination of the viewing angle, and the CO emission in the V-shape is bright at the front side of the collision.

In Table\,\ref{table:1} we see that only the youngest two cases, RCW\,38 and M20, show the bridging feature toward the O star formation. RCW\,38 shows the bridging feature within 0.5\,pc of the $\sim$\,20 O star candidates, where the small colliding cloud is localized \citep{fuk2016a}. The connection among the three features [the bridging feature, $\sim$\,20 O stars, and the small colliding cloud] is tight \citep[see Figure 3 in][]{fuk2016a}. M20 shows three bridging features and one of them is found toward the O star \citep{tor2017}. The three super star clusters, NGC\,3603, Westerlund\,2 and [DBS2003]\,179, show the bridging features in the outer regions of more than 10\,pc radius of the cluster, where molecular gas still remains. RCW\,120 also shows the bridging features outside the H{\sc ii} region in CO and H{\sc i} \citep{tor2015}. N159W-South and N159E-Papillon have a small projected velocity separation of a few\,km\,s$^{-1}$ and the bridging features are not seen.

{
The synthetic observations in Figures \ref{fig3} and \ref{fig4} show that the two velocity components are not clearly separable because of the momentum exchange in the collision which produces intermediate velocity features between the initial two clouds. 
It is therefore not correct to assume that the colliding clouds are always observable as two clearly separate clouds, and observational discrimination of cloud-cloud collision involves intrinsically some ambiguity. 
Figure \ref{fig5} shows that collision is not discernible for $\theta=90$ deg, whereas the probability for the large inclination angle is smaller than 1.5 \% at $\theta$ greater than 80 deg, if the cloud relative motion is randomly distributed. 
Importantly, in spite of these obstacles we are able to use the following three methods as dependable tools to identify cloud-cloud collision as long as the ionization is not significant;
\begin{enumerate}
\item[A)] Complementary distribution between the two colliding clouds: velocity channel distributions are used to identify the two velocity components at the initial two velocities and the two distributions show complementary distribution (Figures \ref{fig3}i and \ref{fig4}i). The two distributions show generally relative displacement depending on the inclination angle of the relative motion to the line of sight (Figure \ref{fig4}i). The intermediate velocity between them, which is mixed up by the interaction, is to be excluded in this comparison.
\item[B)] Position-velocity distribution: the small cloud colliding appears as a V-shaped protrusion including the bridge in a position-velocity diagram along the projected direction of collision, providing a characteristic feature to collision.
\item[C)] Distribution of the first moment: distribution of the first moment serves as another tool to pick out the two components (Figure \ref{figmom1}). This gives a similar distribution to the complementary distribution A), whereas the displacement may be masked in the first moment due to averaging. 
\end{enumerate}
}

We note that the cloud dispersal by the ionization and the collisional merging of the clouds, in addition to the projection effect, can weaken or extinguish these signatures, and one is required to carefully examine these effects which can smear out the collision signatures. Even if a cloud-cloud collision happens, it is possible that one sees none of the signatures. In order to catch the signatures observationally it is required that the system is young, without significant dispersal of the natal gas, and that the projection effect is small. In more than a few\,Myrs after the O star formation by collision, it probably becomes impossible to find any of the observable signatures, for instance, as in case of the relatively young cluster NGC\,6231 with an age of 3\,--\,5\,Myr \citep{bau1999} which has no molecular gas within 30\,pc of the center (Y. Fukui et al. 2018, in preparation).

\section{$^{12}$CO distribution in M42 and M43}
\subsection{$^{12}$CO datasets}
The $^{12}$CO($J$\,=\,1--0) distributions in the M42\,--\,M43 region obtained with the NRO 45-m telescope are presented and discussed by \citet{shi2011}. The observational parameters are listed in Table\,\ref{table:3}. The paper shows that the Orion A molecular cloud is singly peaked in velocity, which is consistent with the previous CO results summarized in Section 1. The cloud properties created in cloud-cloud collision may however show no obvious double component (Section 2), and it is worthwhile to make a thorough investigation of the kinematic properties in the M42\,--\,M43 region from a view point of the cloud-cloud collision scenario. The numerical simulations of the cloud-cloud collision in Section 2 assumed physical parameters which do not exactly coincide with the observed properties of the present cloud. Nonetheless, it is conceivable that the basic characteristics of the colliding clouds hold for different physical parameters, and we here consider the qualitative properties presented in Section 2. In the followings, we apply an X$_{CO}$ factor of 2\,$\times$\,10$^{20}$ (K\,km\,s$^{-1}$)$^{-1}$\,cm$^{-2}$ \citep{str1988} to convert the $^{12}$CO($J$\,=\,1--0) intensity into the molecular column density.

\subsection{The possible two velocity components of $^{12}$CO}
Figure\,\ref{fig6} shows the Decl.-velocity diagram covering $\sim1^\circ$ in Decl. integrated in a R.A. range from $5^\mathrm{h}34\fm6$ to $5^\mathrm{h}35\fm8$, and Figure\,\ref{fig7} shows velocity-channel distributions in a region of $\sim0.5^\circ$ (R.A.) by $\sim1^\circ$ (Decl.) centered on $\theta^{1}$\,Ori. Although most (75\%) of the CO emission is apparently seen in a velocity range of 6\,--\,13\,km\,s$^{-1}$ as a single component, in Figure\,\ref{fig7} we find the distribution of the $^{12}$CO emission is significantly different between the two velocity ranges 5.3\,--\,9.8\,km\,s$^{-1}$ (panels (b), (c), and (d)) and 11.4\,--\,14.4\,km\,s$^{-1}$ (panels (f) and (g)). The strongest peak of the $^{12}$CO emission is seen at 9\,km\,s$^{-1}$ toward $\theta^{1}$\,Ori where the $^{12}$CO linewidth is enhanced from 5\,km\,s$^{-1}$ to 14\,km\,s$^{-1}$ at a 1 K\,km\,s$^{-1}$ level toward the ONC. This main peak having a size of $\sim$\,1\,pc\,$\times$\,1\,pc is called the OMC-1 clump in the followings. We see that the blue-shifted emission at $\sim$\,8\,km\,s$^{-1}$, which constitutes the integral shape of the cloud \citep[mainly in panel (d) of Figure\,\ref{fig7},][]{bal1987}, is extended on the northern and southern sides of the OMC-1 clump, whereas the red-shifted component at $\sim$\,12\,km\,s$^{-1}$ (mainly in panel (f) of Figure\,\ref{fig7}) has a sharp intensity decrease toward the south at Decl.\,=\,$-5\arcdeg\,30\arcmin$. We see the trend also in Figure\,\ref{fig6}. These different spatial distributions at the two velocity ranges suggest a possibility that the two clouds are overlapping as modeled later in Section 4. The intermediate velocity range 10\,--\,11\,km\,s$^{-1}$ (panel (e) of Figure\,\ref{fig7}) shows mixed morphology of the two velocity components.

\subsection{The complementary distributions}
Based on the empirical signatures of cloud-cloud collision (Section 2), we searched the $^{12}$CO data for complementary distribution and found three possible pairs. Figure\,\ref{fig8} shows the three complementary distributions between the blue-shifted and red-shifted clouds in panels (a), (b) and (c).  Figure\,\ref{fig8}(a) shows the distribution toward M42, and Figure\,\ref{fig8}(b) toward M43 and OMC-2.  Figure\,\ref{fig8}(c) is the distribution toward OMC-3, where no H{\sc ii} region is associated \citep[e.g.,][]{pet2008}. 

In Figure\,\ref{fig8}(a), we define the OMC-1 clump in the blue-shifted cloud at the contour level of 15\,K\,km\,s$^{-1}$. We find the U-shaped cloud in part of the red-shifted cloud appears to surround the OMC-1 clump, and define the cloud at the contour level of 2 K\,km\,s$^{-1}$. The defined levels are used for calculating the cloud mass. In Figures\,\ref{fig8}(b) and (c) we find that the red-shifted component shows complementary distributions to the blue-shifted component as detailed below.

In order to show the complementary distributions clearly Figure\,\ref{fig9} overlays the two velocity components in the three regions with appropriate displacements as indicated by arrows in Figures\,\ref{fig9}(b) and (c), whereas no displacement is applied in Figure\,\ref{fig9}(a). In Figure\,\ref{fig9}(a), the complementary distribution is clear in the two components. The OMC-1 clump is extended by more than 1\,pc only toward the north-west at an intensity level of $\sim7$\,K\,km\,s$^{-1}$ (Figure\,\ref{fig8}). It is possible that this extension forms a complementary pair with the intensity depression in the red-shifted component toward (R.A., Decl.)\,=\,($5^\mathrm{h}34^\mathrm{m}40^\mathrm{s}$, $-5\degr15\arcmin$) (panel (f) of Figure\,\ref{fig7}), although the diffuse nature of the extension makes the distribution less evident than toward OMC-1. We note that the intensity depression in the south of the OMC-1 clump in the blue-shifted component, which is extended from R.A.\,=\,$5^\mathrm{h}35\fm1$ to $5^\mathrm{h}35\fm8$ at Decl.\,=\,$-5\degr32\arcmin$, shows also a possible complementary distribution to the red-shifted component in the same direction. Further, in the south of the OMC-1 clump we see an intensity depression elongated in the east to west (R.A.\,=\,$5^\mathrm{h}35^\mathrm{m}00^\mathrm{s}\,$--$\,45^\mathrm{s}$, Decl.\,=\,$-5\degr34\arcmin$\,--\,$30\arcmin$) in the blue-shifted component (Figure\,\ref{fig10}(a)). This depression seems similar in shape to the southern part of the complementary distribution of the red-shifted component, and possibly represents another sign of complementary distribution. The depression appears to be shifted to the south as compared with the red-shifted component. A displacement similar to Figure\,\ref{fig9}(b) may be applicable, while the large size of the OMC-1 clump $\sim$\,1\,pc tends to complicate the complementary correspondence more than in Figure\,\ref{fig9}(b).

In order to explain the estimate of the displacement in Figure\,\ref{fig9}(b), we show in Figure\,\ref{fig10}, the upper panels, the inner part of the two $^{12}$CO components toward the central part of M42 and M43 in the red- and blue-shifted velocity ranges. Figure\,\ref{fig10} also shows enlarged $^{12}$CO images toward M43 in the lower two panels. The blue-shifted cloud shows an intensity depression toward the exciting star NU\,Ori, which is elongated in the north-south by $\sim$\,0.5\,pc with a width of  $\sim$\,0.2\,pc as enclosed by a box in Figure\,\ref{fig10}(c). The eastern side of the depression is bound by a filamentary feature that corresponds to the dark lane in M43 (see also Figure\,\ref{fig12}). A corresponding emission feature to the intensity depression is seen in the red-shifted component which has a similar shape to the depression. We call the two features as the $``$Orion Keyhole$"$ and the $``$Orion Key$"$ in order to avoid confusion with the Keyhole Nebula in Carina. We applied the method to estimate the displacement in the complementary distribution which is presented in the Appendix. In fitting the displacement, we restricted the area of the analysis to the region of the Orion Keyhole and the Orion Key as shown in Figure\,\ref{fig11}. By inspection we find that the Orion Key has a discontinuous $``$bent$"$ in the middle at $\sim$\,0.25\,pc from the southern edge, which seems to morphologically correspond to the edge of the eastern filament of the Orion Keyhole (Figure\,\ref{fig8}).  We then chose the direction of the displacement at a position angle of $108^\circ$ so that the two features coincide spatially after the displacement. Figure\,\ref{fig11} shows the overlapping function $H(\Delta)$ in pc$^{2}$ defined in the Appendix and indicates that $\Delta$\,=\,0.3\,pc gives the optimum fit. We confirmed that the present position angle adopted gives the best fit, i.e., the maximum value of $H(\Delta)$, by changing the angle in a range $103^\circ$\,--\,$113^\circ$ A displacement of 0.3\,pc shown by the arrow in Figure\,\ref{fig9}(b) gives the fit between the Orion Keyhole and Key. By the displacement, we also see that the western edge of the red-shifted component shows good correspondence with the eastern edge of the blue-shifted component over $\sim$\,1\,pc in the north-south, which lends support for the displacement {(see dashed green line in Figure\,\ref{fig9}(b))}.

In Figure\,\ref{fig9}(c), while not so clear as in the other two cases, the two peaks of the blue-shifted component at (R.A., Dec)\,=\,($5^\mathrm{h}35^\mathrm{m}10^\mathrm{s}$, $-5\degr00\arcmin$) and ($5^\mathrm{h}35^\mathrm{m}15^\mathrm{s}$, $-5\degr05\arcmin$) are located toward the intensity depression of the red-shifted component, forming a possible pair of complementary distribution, after a displacement of 0.1\,pc shown by an arrow. This displacement was chosen based on eye inspection because the distribution is relatively simple. 

\subsection{{The bridging features}}
{Figure\,\ref{fig:pv_abc} shows the position-velocity diagrams of the three regions in Figures\,\ref{fig8}(a)--(c), where the integration ranges are indicated as filled white areas in Figure\,\ref{fig8}. While in Figures\,\ref{fig:pv_abc}(a) and (c) the CO velocity distributions along R.A. are presented, the CO distribution along the X-axis defined in Figure\,\ref{fig11}(a) is shown in Figure\,\ref{fig:pv_abc}(b). In Figure\,\ref{fig:pv_abc}(a), which includes the U-shaped cloud in the red-shifted cloud and the OMC-1 clump in the blue-shifted cloud, the CO emission shows a V-shaped distribution as depicted by dashed lines. V-shaped gas distribution in the position-velocity diagram is an observational signature of cloud-cloud collision as discussed in Section\,2.4, and the V-shape in Figure\,\ref{fig:pv_abc}(a) resembles the position-velocity diagram of the synthetic CO data in the $\theta=0^\circ$ case in Figure\,\ref{fig3}(k). In Figures\,\ref{fig:pv_abc}(b) and (c), toward M43 and OMC-3, respectively, the CO emission shows V-shaped velocity distribution with skew, which are similar with the $\theta=45^\circ$ case in Figure\,\ref{fig4}(k) than the $\theta=0^\circ$ case. This is consistent with the complementary distributions with displacement in Figures\,\ref{fig9}(b) and (c).}

\subsection{{Comparisons with the optical image and the first moment}}
Figure\,\ref{fig12} show overlays of the two velocity components on the optical image. The most notable correspondence is seen toward M43 where the thin dark lane in the center of the nebula coincides with the CO filament in the east of the Orion Keyhole (Figure\,\ref{fig12}(a)). The other blue-shifted features of $\sim$\,0.5\,deg in length show coincidence with the dark areas in the northeast of M42, and the visual extinction of this direction is high ($A_{\rm v} \gtrsim 30$: \citet{Scandariato2011}). 
This correspondence is consistent with that the blue-shifted component separated by 1--4 pc from the core of the O stars lies on the nearside of the optical nebula. 
On the other hand, it is well known that the visual extinction toward the Trapezium stars is small ($A_{\rm v} \lesssim 3$: \citet{Scandariato2011}). 
Therefore, we suggest that the OMC-1 clump as a part of blue-shifted cloud is on the far side of the optical nebula. This reflects the 3-dimensional distribution of the clouds (\citet{Balick1974}), and is consistent with the Champaign flow model (\citet{tenorio1979}).
The red-shifted component in Figure\,\ref{fig12}(b) shows no correspondence with the optical features, suggesting that the red-shifted component lies on the far side or the inside of the nebula. We remark that the Orion Key has no corresponding dark lane in M43, which is consistent with the present collision. 

Figure \ref{fig-ori-mom1} shows the distribution of the first moment overlayed on the blue-shifted and red-shifted components in contours. 
Generally, the first moment distribution shows good correspondence; the blue-shifted component corresponds to the first moment at velocity smaller than 10 km s$^{-1}$, and the red-shifted component to the Orion Key and the U shaped cloud. 
This lends additional support to the present interpretation of the two components. 

\section{The cloud-cloud collision model}
\subsection{Model}
We present a model of the two molecular clouds which are colliding to trigger the formation of the O\,/\,B stars in M42 and M43. Table\,\ref{table:5} lists the main physical parameters of the model clouds and Figure\,\ref{fig14} gives a schematic of the clouds before and after the collision. The model consists of two clouds of projected velocities peaked at $\sim$\,8\,km\,s$^{-1}$ and $\sim$\,13\,km\,s$^{-1}$, while the actual relative velocity between the two clouds is in the order of 10\,km\,s$^{-1}$. Both of the clouds are elongated along the Galactic plane. The mass of the blue-shifted component is dominant ($\sim$80\%) toward OMC-1, explaining the apparent single peak, while the border velocity 11\,km\,s$^{-1}$ between the two model clouds may have some uncertainty. 
The intermediate velocity gas is not clearly separable in velocity because of the gas mixing in the collision (e.g., Takahira et al. 2014). It is difficult to separate the compressed layer clearly in the intermediate velocity as shown theoretically by Figures \,\ref{fig4} and \,\ref{fig5} and as confirmed observationally by Figure \,\ref{fig14}. We therefore gave the velocity ranges of the two components at a representative boundary of 11 km s$^{-1}$ in Table\,\ref{table:5}.
The total mass of the two clouds  $\sim$\,10$^{4}$\,$M_\odot$ is too small by an order of magnitude to gravitationally bind the estimated relative velocity $\sim$\,7\,km\,s$^{-1}$. The model clouds have an overlap toward the OMC-1 clump at Decl.\,=\,$-5\degr30\arcmin$\,--\,$-5\degr18\arcmin$ which corresponds to the region of the enhanced velocity span in Figure\,\ref{fig6}. The enhancement is interpreted as due to the turbulence by the collisional interaction \citep{ino2013}, while the effect of the protostellar outflow may also be in part responsible for the enhancement only in the small region close to the protostar. The model offers an explanation of the complementary distribution which is found in the present clouds. The collisional interaction is seen at least in the three regions as indicated by the complementary spatial distribution; one is toward M42 and triggered formation of $\theta^{1}$\,Ori and $\theta^{2}$\,Ori (Figure\,\ref{fig9}(c)), and another toward M43, which triggered formation of NU\,Ori (Figure\,\ref{fig9}(b)). The other is toward the north of M43 without an H{\sc ii} region (Figure\,\ref{fig9}(a)). Two more regions associated with possible complementary distribution are noted in Section 3, while no clear star formation is seen except for M42 and M43. In the present case, the half-power linewidths of the two clouds are 2\,--\,3\,km\,s$^{-1}$ and the projected velocity separation $\sim$\,5\,km\,s$^{-1}$ is too small for separating the possible bridging feature.

In the model, $\theta^{1}$\,Ori and $\theta^{2}$\,Ori, including nearly ten O stars, were formed on the near side of the blue-shifted cloud which was collided and shock-compressed by the more extended red-shifted cloud. This collision created the cavity in the red-shifted cloud following the cloud-cloud collision model \citep{hab1992, ana2010, tak2014}. The interface layer between the two clouds became strongly-compressed and highly turbulent to form O stars as shown by \citet{ino2013}. The O stars formed on the nearside of the dense blue-shifted cloud ionized the nearside of the cloud to form the M42 Nebula as exposed to us. NU\,Ori was formed in the collision between the Orion Key and the red-shifted component having the Orion Keyhole. NU\,Ori is a B3 star and ionized M43 (see Section 4.3 for more details). The ionization possibly has widened the size of the Orion Keyhole, whereas the smaller UV photon flux of NU\,Ori than $\theta^{1}$\,Ori and $\theta^{2}$\,Ori is not powerful enough to dissipate the Orion Keyhole at present. 

\subsection{Timescales and the collision}
The timescale of the collision is approximately estimated from a ratio of the cloud size and the relative velocity between the two clouds. If we assume tentatively the relative motion between the two clouds has an angle of $45^\circ$ to the line of sight, the relative velocity corrected for the projection is estimated to be $\sim$\,7\,km\,s$^{-1}$, which is consistent with the enhanced $^{12}$CO broadening toward the ONC (Figure\,\ref{fig6}). The apparent spatial extent of the colliding clouds forming O\,/\,B stars ranges from $\sim$\,0.3\,pc to $\sim$\,0.8\,pc (Figure\,\ref{fig8}). The ratio between the timescale and the cloud size yields roughly 0.05\,--\,0.11\,Myr as the collision timescale and poses a constraint on the upper limit for the age of the stars formed by triggering.

The complementary distribution is characterized by the projected displacement between the colliding clouds (Figure\,\ref{fig3}, see also the Appendix). The displacement is small for a small collision timescale. Also, the displacement $relative$ to the size of the small cloud becomes smaller, if the size of the small cloud in the collision model is large. Figure\,\ref{fig10} shows different displacements between the two velocity components among the three regions. We apply a detailed analysis of the timescale to the Orion Key and Keyhole, where a displacement is obtained on their relative location by the analysis in Section 3 thanks to the small size, $\sim$\,0.2\,pc, of the small cloud. The separation between the two features are estimated to be 0.3\,pc at a relative velocity is 4\,km\,s$^{-1}$. If we assume 45\,deg as the angle of the relative motion to the line of sight, the time scale of the collision is estimated to be 0.4\,pc\,/\,5.7\,km\,s$^{-1}$\,=\,0.07\,Myr if no deceleration of the relative motion is assumed. The northern region in Figure\,\ref{fig9}(c) shows a displacement of $\sim$\,0.1\,pc, while it may be less accurate than in Figure\,\ref{fig9}(b). The OMC-1 clump in Figure\,\ref{fig9}(a) shows no appreciable displacement, whereas the larger clump size, $\sim$\,1\,pc, tends to smear out a possible displacement. The separation between the two velocity components may vary from region to region reflecting the initial distribution of the two components, leading to different collision parameters. In summary, we conclude that the typical collision timescale is $\sim$\,0.1\,Myr with some regional minor variation.

\subsection{The star formation under triggering}
At the present epoch, the blue-shifted component has high peak column density toward the peak of the OMC-1 clump, $\sim$\,2\,$\times$\,10$^{23}$\,cm$^{-2}$, while the red-shifted cloud has significantly smaller column density, $\sim$\,10$^{22}$\,cm$^{-2}$. In M43, the typical column density in the Orion Key and Keyhole is $\sim$\,10$^{22}$\,cm$^{-2}$. We assume that the initial conditions of the two clouds are similar to those of the present clouds, whereas it is probable that the present column densities have become somewhat lower than the initial values due to the collisional interaction and ionization by the formed stars. By comparing O star formation in several regions, \citet{fuk2016a} argued that the number of O stars formed by collisional triggering depends on the initial column density; more than ten O stars are formed for a threshold column density $\sim$\,10$^{23}$\,cm$^{-2}$ and a single O star for a threshold column density $\sim$\,10$^{22}$\,cm$^{-2}$. The difference in the O\,/\,B star formation in M42 and M43 is consistent with the suggestion.

The MHD numerical simulations by \citet{ino2013} have shown that the interface layer between the colliding clouds become denser and more turbulent with amplified magnetic field, realizing the high mass accretion rate $\sim$\,10$^{-4}$\,--\,10$^{-3}$\,$M_\odot$\,yr$^{-1}$ which satisfies the formation of the high-mass stars greater than 20\,$M_\odot$. In the present typical collision timescale 0.1\,Myr, the stellar mass which is attained by mass accretion is 10\,--\,100\,$M_\odot$. So, the present short timescale is consistent with the O star formation.

More details of the star formation may be explored. In M42, the ionized gas has higher density in $\theta^{1}$\,Ori than in $\theta^{2}$\,Ori. There are even younger heavily-embedded high-mass protostars showing outflows, the BN\,/\,KL object and the Orion-S region (e.g., \cite{dra1983}), in the west of $\theta^{1}$\,Ori. We see an age sequence from the east to the west, where $\theta^{2}$\,Ori is the oldest and the BN\,/\,KL object and the Orion-S region the youngest. A possible scenario is that the age sequence is caused by the time sequence of the collision; i.e., the colliding two components has a spatial gradient in the sense that the western part collided first and then the collision is propagating to the west, and the most recent triggering formed the BN\,/\,KL object and the Orion-S region. 
 A similar age gradient is seen in RCW\,38; the eastern part is already ionized, whereas the western part is more heavily embedded \citep{fuk2016a}.

In M43, the position of NU\,Ori is separated to the west by $\sim$\,0.15\,pc from the eastern filament of the Orion Keyhole. The velocity of the interface layer generally becomes smaller by the momentum conservation \citep[e.g.,][]{haw2015b}, which places the formed star behind the small cloud. A possible scenario is that the B star formation took place by the collision in the eastern edge of the Orion Key. This explains the location of NU\,Ori and the situation is similar to M20 where a single O star is formed by triggering in the eastern edges of the clouds \citep{tor2017}. 

\subsection{The age of the ONC}
The age of the ONC members ranges from less than 0.1\,Myr to more than 1\,Myr while the majority is in the range 0.1\,--\,1\,Myr as estimated in the HR diagram by comparing with the theoretical stellar evolutionary tracks \citep{hil1997}. The collision timescale less than 0.1\,Myr corresponds to the youngest end of the age distribution. We infer that the blue-shifted cloud which dominates the molecular column density in the Orion A cloud was forming low-mass stars of the ONC prior to the collision in the last  1\,Myr. The distribution of low-mass young stars of the ONC, 2MASS variables stars in Figure\,\ref{fig15} \citep{car2001}, indicates that these stars toward the OMC-1 clump are correlated well with the column density distribution of the blue-shifted cloud except for the $^{12}$CO peak including the BN\,/\,KL object where the extinction may be large at K band \citep{fei2005}. 
{
In addition, it is possible that the 2MASS variable stars are obscured by several  magnitude in K band toward the molecular peak of the OMC-1 clump (R.A.$\sim 5^\mathrm{h}35\fm3$ and Decl.$\sim -5 \degr 20\arcmin$--$-5\degr30\arcmin$) as suggested by their decrease in number by $\sim20$ \% (Figure \ref{fig15}b). 
}
These low-mass members are spatially more extended than the O stars \citep{hil1997}, beyond the region of the collisional interaction. We see little molecular gas in the two regions in the north of M42, (R.A., Decl.)\,=\,($5^\mathrm{h}35\fm4$\,--\,$5^\mathrm{h}36\fm0$, $-5\degr15\arcmin$\,--\,$-5\degr00\arcmin$), and the south of the M42 clump (R.A., Decl.)\,=\,($5^\mathrm{h}35\fm0$\,--\,$5^\mathrm{h}35\fm7$, $-5\degr28\arcmin$\,--\,$-5\degr32\arcmin$), whereas 2MASS stars are distributed there. The two correspond to the regions of the suggested collisional interaction and the blue-shifted gas in the regions has possibly been removed by the collision after the low-mass star formation.  The red-shifted cloud had lower column density, not actively forming stars before the collision, while the northern part of the red-shifted component may have been forming part of the stars as indicated by the spatial correspondence with the stars (Figure\,\ref{fig15}). It is possible that not only O\,/\,B stars but also low mass-stars were formed by the collisional triggering \citep[cf.][]{nak2012}. The young low-mass stars in the HR diagram by \citet{hil1997} suggest very young low-mass stars of an age less than 0.1\,Myr. Part of them might represent protostars whose accretion disk was destructed by the UV photons of $\theta^{1}$\,Ori and $\theta^{2}$\,Ori before they reach $>$\,10\,$M_\odot$.

\subsection{{Alternative interpretations}}

{
H{\sc ii} regions containing high temperature ionized gas are expanding and H{\sc ii} expansion can be an alternative to explain the gas motion instead of cloud-cloud collision. 
\cite{sug1986} examined $^{13}$CO($J$\,=\,1--0) emission over the M42/M43 region in order to test the gas motion driven by the H{\sc ii} region.
These authors found it more likely that the gas motion is dominated by the pre-existent motion unrelated to the H{\sc ii} expansion. 
Figure \ref{fig:pv_abc}c shows that the velocity span of the $^{12}$CO emission is not particularly enhanced toward the peak of the H{\sc ii} region, where one expects most significant acceleration if H{\sc ii} expansion plays a role. 
Conversely, the span is most enhanced at the eastern edge of the molecular gas with no systematic trend. 
Figure \ref{fig6}, another position velocity cut in Decl., shows that the velocity span is nearly uniform at $\sim$ 8 km s$^{-1}$ at a contour level of $\sim$ 2 K degree over 1 pc. 
This is the OMC-1 clump and is part of the blue-shifted component. 
The clump is on the far side of the O stars and the blue shift is in the opposite sense to the acceleration by H{\sc ii} expansion. 
Figure \ref{fig:pv_abc}b, a position velocity cut in M43, also shows that the velocity span has no enhancement toward NU Ori except for the Orion Key in a small size of 0.1 pc. 
These current kinematic signatures are consistent with the conclusion reached by \cite{sug1986}.
In other H{\sc ii} regions, significant molecular gas motion driven by H{\sc ii} regions is not reported (e.g., RCW38, \cite{fuk2016a}; RCW120, \cite{tor2015}; NGC6334, \cite{2018arXiv170605771F}; M20, \cite{tor2017}; M16, \cite{2018arXiv170606002N}; M17, \cite{2018arXiv170606956N}). 
In RCW38, where two velocity components are observed, the red-shifted gas is on the nearside of the H{\sc ii} region, supporting that H{\sc ii} expansion does not dominate gas motion, and in RCW120, a typical Spitzer bubble, $^{12}$CO observations found no sing of expansion due to the H{\sc ii} region. 
Theoretical studies of H{\sc ii} expansion (e.g., \cite{hos2005}) showed that the molecular gas layer accelerated by H{\sc ii} expansion is thin, in the order of 0.1 pc, and that the bulk of the molecular gas cannot be accelerated prior to ionization. 
The observations above are consistent with these theoretical works. 
Alternatively, H{\sc ii} gas may be pressurized and accelerate ambient molecular gas unionized. 
The molecular gas is however always clumpy, and it is unlikely that H{\sc ii} region is totally confined by the molecular gas. 
This suggests that H{\sc ii} gas escapes easily through low-density holes without molecular acceleration. 
It is more likely that high mass stars accelerate surrounding gas via supernova explosion as observed in molecular super shells formed by multiple supernova explosions over a larger timescale of $\sim$20 Myr (e.g., \cite{fuk1999}; \cite{mat2001}; \cite{daw2008}). 
Such acceleration may explain the origin of the supersonic gas motion which leads to cloud-cloud collision over a timescale larger than 100 Myr.
}

\section{Discussion}
\subsection{Theories of high-mass star formation under gravitational binding}
The monolithic collapse and the competitive accretion are the two scenarios for massive cluster formation and have been discussed for more than 
a decade \citep[e.g.,][]{zin2007}. In either case the massive gas/star system is self-gravitationally bound and multiple sources evolve into stars of various masses by mass accretion. It is also discussed that the bound system may evolve in time and becomes unbound by gas dispersal \citep[e.g.,][]{kro2001etal}. These two gravitationally-bound scenarios have no explicit physical mechanism of realizing the high-mass accretion rate like the shock-induced turbulence in the cloud-cloud collision scenario, and it is assumed that the gas achieves the high-mass accretion rate by some gravitationally driven mechanism. The colliding clouds in the present hypothesis are not gravitationally bound by the supersonic velocity and the collision increases the mass accretion rate to be $10^{-4}$\,--\,$10^{-3}$\,$M_\odot$\,yr$^{-1}$, which is higher than in the bound scenarios, favoring the high-mass star formation \citep{ino2013}. 

\subsection{Evolutionary timescales and mass segregation}
The evolution of the monolithic mass aggregation has been tested theoretically. In the simulations of proto-clusters, a single gravitationally bound N-body system is adopted as the initial condition of a cluster whose number of stars are assumed to be 800 -- 2000 in the same order with that of the ONC \citep[e.g.,][]{kro2001etal, ban2015}. The effect of the stellar feedback is also incorporated in recent simulations \citep[e.g.,][]{dal2013}.

The typical timescale of the evolution of the N-body system is Myrs, which is significantly larger than that of the cloud-cloud collision，less than 0.1\,Myr. It is probable that these simulations are applicable to the pre-collision cluster of low-mass stars in the Orion blue-shifted cloud of the present hypothesis. Some N-body simulations assume $``$rapid$"$ gas dispersal by ionization due to the high-mass stars, and it is argued that such rapid dispersal is consistent with the observations  \citep{ban2015}. The rapid dispersal is however yet of a large time scale, close to Myrs. There is increasing observational evidence for a short time scale less than 0.1\,Myr in O star formation; observations of the ONC shows that the youngest stars have age less than 0.1\,Myr, and in the two super star clusters NGC\,3603 and Westerlund\,1 the age spread of young clusters is short in the order of 0.1\,Myr \citep{kud2012}. It is also becoming established that the age and mass segregation is outstanding in young massive clusters; the O stars are concentrated in the inner most part of NGC\,3603, the ONC, and R\,136. While the stellar mass function depends on the area for averaging the stellar properties, the innermost part tends to have a top-heavy mass function in these clusters. We need to be cautious that the ONC may consist of at least two different populations. The ONC has a usual mass function similar to the field initial mass function IMF having a Salpeter-like slope \citep{kro2001}. In the present model the stars toward M42 is a superposition of two different populations; the older low-mass members formed prior to the collision are likely dominant in the blue-shifted cloud and the other youngest stars including O\,/\,early B stars were formed in triggering by the collision, which are not dominant in mass. The pre-existent low-mass stars dominate the stars of the ONC, making the steep slope of the mass function, while about a half of them may still be deeply embedded in the blue-shifted cloud \citep{hil2000, mue2008}. One may expect that the IMF has discontinuous slopes if the star formation by the two different modes are working. We suggest that the total mass of the O stars in the center of the ONC is $\sim$\,150\,$M_\odot$ \citep{mue2008}. On the other hand, the molecular mass of the OMC-1 clump is 3.3\,$\times$\,10$^{3}$\,$M_\odot$. If we assume tentatively that the depth where of the shock front propagated in the OMC-1 clump is 0.3\,pc, one third of the clump size, the molecular mass already shocked is estimated to be $\sim$\,1000\,$M_\odot$, large enough to form the ten O stars of 20\,--\,30\,$M_\odot$. The interface layer is included in these two components for the collisional area of $\sim 1$ pc$^{2}$ toward theta1 and $\theta_2$ Ori, and NU Ori, although the layer is not clearly separable in velocity because of the turbulent mixing by the collision. The present collision model suggests that the forming stars by triggering are located on the front side of the M42 clump, producing a non-spherically symmetric distribution of the cluster. This is a natural consequence of the collision that has a directivity at least in the early phase where part of the molecular gas still remains without ionization. We note that the N-body simulations result in spherically symmetric system after Myrs do not explain such asymmetry. The total mass of the Orion Key and Keyhole region is estimated to be $\sim$150\,$M_\odot$ for a size of $\sim$\,0.5\,pc. This is large enough to form NU\,Ori with 17\,$M_\odot$ \citep{mue2008}.

The initial conditions of the present cloud-cloud collision assume a molecular cloud having high column density 10$^{23}$\,cm$^{-2}$, which has been continuously forming low-mass stars. These low-mass stars are not able to ionize the cloud and the blue-shifted cloud continues to form low-mass stars until it forms high-mass stars which are capable of ionizing the cloud. The O star formation is suddenly triggered by the impact of the supersonic collision with the red-shifted cloud which has lower column density. The epoch of O star formation may range in a certain time span, whereas the cloud-cloud collision poses a limit on the age spread of the youngest members including O stars by the ratio of the cluster size divided by the collision velocity to be $\sim$\,0.1\,Myr. It is unlikely that the two gravitationally bound scenarios can explain the very short discontinuous history of star formation less than 0.1\,Myr. We do not see any gap in the age distribution of the cluster member. This is a natural consequence of the cloud-cloud collision following\ Myr low-mass star formation.

\section{Conclusions}
We have made a new analysis of the molecular gas toward M42 and M43, and test a hypothesis that two clouds collided about 0.1\,Myr ago to form the northern part of the Orion A cloud. We have shown the collision model offers a reasonable explanation of the observed cloud and cluster properties in the M42 and M 43 region. Although, currently, we do not have an extensive observational basis that the cloud-cloud collision is a dominant mechanism of O star formation in the Galaxy. The main conclusions of the present study are summarized as follows.
\begin{enumerate}
\item[1.] The present analysis of the high resolution CO data shows a possibility that the Orion A cloud toward M42 and M43 consists of two velocity components, one at 4.0\,--\,11.1\,km\,s$^{-1}$ and the other at 11.1\,--\,14.9\,km\,s$^{-1}$. The mass of the blue-shifted component is 1.5\,$\times$\,10$^{4}$\,$M_\odot$ with its velocity range 4.0\,--\,11.1\,km\,s$^{-1}$ and the red-shifted component 3.4\,$\times$\,10$^{3}$\,$M_\odot$ with its velocity range 11.1\,--\,14.9\,km\,s$^{-1}$ having a size of 4\,pc by 7\,pc in R.A. and Decl. The interface layer having $\sim 1000\ M_{\odot}$ is included in these two components for the collisional area of $\sim 1$ pc$^{2}$ toward theta1 and $\theta_2$ Ori, and NU Ori, although the layer is not clearly separable in velocity because of the turbulent mixing by the collision.

\item[2.]We find at least three pairs of complementary distributions between the two clouds having scale sizes of 0.1\,--\,1\,pc. These complementary distributions of molecular gas are an observational signature characteristic to cloud-cloud collision according to the hydrodynamical numerical simulations at the early phase of O star formation. The major complementary distribution is found toward $\theta^{1}$\,Ori and $\theta^{2}$\,Ori where the blue-shifted component of $\sim$\,0.5\,pc radius, the OMC-1 clump, is surrounded by the red-shifted component, the U-shaped cloud, which is extended with a radius of $\sim$\,1\,pc. The secondary complementary distribution is seen toward M43 as named the Orion Key and Keyhole, where a spatial displacement of 0.3\,pc is seen between the two complementary distributions. 

\item[3.]We frame a hypothesis that the two velocity components collided with each other to trigger formation of the nearly ten O stars in M42 and the B star in M43. The displacement between the two complementary features in M43 suggests that the relative motion makes a large angle to the line-of-sight. By assuming the angle to be 45\,deg, the relative velocity and separation between the clouds is estimated to be $\sim$\,6\,km\,s$^{-1}$ and $\sim$\,0.5\,pc, respectively, after correction for the projection, and the typical collision time scale is calculated to be $\sim$\,0.1\,Myr. 

\item[4.]Although the collision is extended over an area of 4\,pc\,$\times$\,7\,pc in R.A. and Decl. the triggering of the O\,/\,B star formation is seen only toward the localized regions with OB stars.  We suggest that this trend is consistent with the threshold for O star formation in cloud-cloud collision suggested by \citet{fuk2016a}; multiple O star formation is possible for column density $\sim$\,10$^{23}$\,cm$^{-2}$ while the single O\,/\,B star formation for $\sim$\,10$^{22}$\,cm$^{-2}$. The high segregation of the high-mass cluster members to the cluster center is explained as a reflection of the initial column density distribution prior to the collision. 

\item[5.]The stellar contents of the ONC has been best studied among the rich clusters thanks to its unrivaled small distance to the sun although the cluster is not overwhelmingly rich in high-mass stars. The OB stars in the ONC are minor members in terms of stellar mass, and the initial mass function is not significantly influenced by the formation of the OB stars. The youngest members including the O\,/\,early B stars are formed by the collision within 0.1\,Myr, and the large age spread of the cluster, from a few 0.01\,Myr to a few times 1.0\,Myr, is understood by a combination of low-mass star formation in the blue-shifted cloud prior to the collision and the recent O\,/\,B star formation by the collisional triggering. Formation of O stars may still be continuing toward the western edge of the M42 clump where the BN object and OMC-1-S are located along the north-south direction. 
\end{enumerate}

In the present study we have shown that cloud-cloud collision offers a possible consistent interpretation of the formation of the O\,/\,B stars in M42 and M43. Since the immediate vicinity of $\theta^{1}$\,Ori and $\theta^{2}$\,Ori is significantly ionized, we are not allowed to directly witness the triggering as in case of RCW\,38. We note that the BN\,/\,KL object can be such a candidate as discussed in Section 5, and suggest that a future more careful scrutiny of the region and its surroundings at higher resolutions with ALMA etc. will offer a further insight into the possible role of the collisional interaction.

\acknowledgments
We are grateful to Thomas Haworth for valuable discussion on this paper.
We also thanks to Robert Gendler for optical image of the ONC, and Kazuki Tokuda for ALMA archive data.
This publication makes use of data products from the Two Micron All Sky Survey, which is a joint project of the University of Massachusetts and the Infrared Processing and Analysis Center/California Institute of Technology, funded by the National Aeronautics and Space Administration and the National Science Foundation.
 Nobeyama Radio Observatory is a branch of the National Astronomical Observatory of Japan, National Institutes of Natural Sciences.
The ASTE telescope is operated by National Astronomical Observatory of Japan (NAOJ).
This paper makes use of the following ALMA data: ADS/JAO.ALMA\#2012.1.00554.S. 
ALMA is a partnership of ESO (representing its member states), NSF (USA) and NINS (Japan), together with NRC (Canada), NSC and ASIAA (Taiwan), and KASI (Republic of Korea), in cooperation with the Republic of Chile. 
The Joint ALMA Observatory is operated by ESO, AUI/NRAO and NAOJ.
This work was supported by JSPS KAKENHI grant numbers 15H05694, 15K17607, 15K05014, 16K17664, and 26247026; by JSPS and by the Mitsubishi Foundation.


\clearpage


\begin{table*}
\rotatebox{90}{\begin{minipage}{\textheight}
\centering
\caption{Super star clusters and single O stars formed by cloud-cloud collision and the Orion Nebula Cluster.} 
\begin{tabular}{cccccccccccc}
\hline
\hline
Object & Molecular mass & Molecular & Relative & complementary  & Bridging & Cluster age & Number of & Reference \\
   &  &  column density & velocity & distribution & feature &  & O stars &  \\
             & \ [$M_\odot$] & \ [cm$^{-2}$] & \ [km\,s$^{-1}$] &   &  & \ [Myr] &   &  \\
   \ (1)   & \ (2) & \ (3) & \ (4) & \ (5)  & \ (6) & \ (7) & \ (8) & \ (9) \\
\hline
RCW\,38 & (2\,$\times$\,10$^{4}$, 3\,$\times$\,10$^{3}$) &  (1\,$\times$\,10$^{23}$, 1\,$\times$\,10$^{22}$) & 12 & no & yes & $\sim$\,0.1 & $\sim$\,20 & [1] \\
NGC\,3603 & (7\,$\times$\,10$^{4}$, 1\,$\times$\,10$^{4}$) &  (1\,$\times$\,10$^{23}$, 1\,$\times$\,10$^{22}$) & 15 & no & yes & $\sim$\,2.0 &  $\sim$\,30 & [2] \\
Westerlund\,2 &  (9\,$\times$\,10$^{4}$, 8\,$\times$\,10$^{4}$) &  (2\,$\times$\,10$^{23}$, 2\,$\times$\,10$^{22}$) & 16 & yes & yes & $\sim$\,2.0  & 14 & [3, 4] \\
\ [DBS2003]\,179 & (2\,$\times$\,10$^{5}$, 2\,$\times$\,10$^{5}$) & (8\,$\times$\,10$^{22}$, 5\,$\times$\,10$^{22}$) &  20 & yes & yes & $\sim$\,5.0 &  $>$ 10 & [5] \\
\hline
ONC (M42) &  (2\,$\times$\,10$^{4}$, 3\,$\times$\,10$^{3}$) & (2\,$\times$\,10$^{23}$, 2\,$\times$\,10$^{22}$) & $\sim$\,7$^{(a)}$ & yes & no & $\sim$\,0.1 & $\sim$\,10 & [6] \\
ONC (M43) &  (3\,$\times$\,10$^{2}$, 2\,$\times$\,10$^{2}$) & (6\,$\times$\,10$^{22}$, 2\,$\times$10$^{22}$) & $\sim$\,7$^{(a)}$ & yes & no & $\sim$\,0.1 & $\sim$\,1 & [6] \\
\hline
M20 & (1\,$\times$\,10$^{3}$, 1\,$\times$\,10$^{3}$) &  (1\,$\times$\,10$^{22}$, 1\,$\times$\,10$^{22}$) & 7.5 & yes & yes & $\sim$\,0.3  & 1 & [7] \\
\multirow{2}{*}{RCW\,120} & \multirow{2}{*}{(5\,$\times$\,10$^{4}$, 4\,$\times$\,10$^{3}$)} & \multirow{2}{*}{(3\,$\times$\,10$^{22}$, 8\,$\times$\,10$^{21}$)} &  \multirow{2}{*}{20} & \multirow{2}{*}{yes} & \multirow{2}{*}{yes} & $\sim$\,0.2 & \multirow{2}{*}{1} & [8] \\
&  &  &  &  &  & $<$ 5.0 &  & \ [9] \\
N159W-South & (9\,$\times$\,10$^{3}$, 6\,$\times$\,10$^{3}$) &  (1\,$\times$\,10$^{23}$, 1\,$\times$\,10$^{23}$)  & $\sim$\,8$^{(b)}$ & no  & no & $\sim$\,0.06 & 1 & [10] \\
N159E-Papillon & \scriptsize(5\,$\times$\,10$^{3}$, 7\,$\times$\,10$^{3}$, 8\,$\times$\,10$^{3}$) &  \scriptsize(4\,$\times$\,10$^{22}$, 4\,$\times$\,10$^{22}$, 6\,$\times$\,10$^{22}$) & $\sim$\,9$^{(c)}$ & no & no & $\sim$\,0.2 & 1 & [11] \\
\hline
\hline
\end{tabular} \label{table:1}
\tablecomments{Column: (1) Name. (2, 3) Molecular masses and column densities of the two$\slash$three clouds. (4) Relative velocity between the colliding clouds. (5) complementary distribution towards the center cluster/O-star. (6) Bridging feature between the two clouds. (7, 8) The age and the number of the cluster/O-star. (9) References: [1]\citet{fuk2016a}, [2]\citet{fuk2014}, [3]\citet{fur2009}, [4]\citet{oha2010}, [5] Kuwahara et al., in preparation, [6] The present study, [7]\citet{tor2011}, [8]\citet{mar2010}, [9]\citet{tor2015}, [10]\citet{fuk2015b}, [11]\citet{sai2017}. (a)--(c) corrected for the projection.}
\end{minipage}
}
\end{table*}

\begin{table}[]
\centering
\begin{center}
  \caption{Velocity range of the molecular emission in the 8 cases (Table 1).}
   \begin{tabular}{cccccccccccc}
\hline
\hline
     Object & Velocity range & Reference \\
      & \ [km\,s$^{-1}$] & \\
      (1) & (2) & (3) \\
    \hline
    RCW\,38 & $+3.0$\,--\,$+34.0$ & [1]\\
    NGC\,3603 & $+1.2$\,--\,$+20.9$ & [2] \\
    Westerlund\,2 & $+1.2$\,--\,$+8.7$ & [3, 4] \\
    \ [DBS2003]\,179 & $-104.0$\,--\,$-81.0$ & [5] \\
    \hline
    ONC & $+4.0$\,--\,$+14.9$ & [6] \\
    \hline
    M20 & $-1.0$\,--\,$+30.0$ & [7] \\
    RCW\,120 & $-40.0$\,--\,$+8.8$  & [8] \\
    N159W-South & $+200.0$\,--\,$+269.8$ & [9] \\
    N159E-Papillon & $+200.0$\,--\,$+269.8$ & [10] \\
\hline
\hline
\end{tabular} \label{table:2}
\end{center}
\tablecomments{Column: (1)object name. (2)velocity range. (3)References: [1]\citet{fuk2016a}, [2]\citet{fuk2014}, [3]\citet{fur2009}, [4]\citet{oha2010}, [5] Kuwahara et al., in preparation, [6]The present study, [7]\citet{tor2011}, [8]\citet{tor2015}, [9]\citet{fuk2015b}, [10]\citet{sai2017}.}
\end{table}

\begin{table}[]
\begin{center}
  \caption{The initial conditions of the numerical simulations \citep{tak2014}}
   \begin{tabular}{cccccccccccc}
    \hline    \hline
    Box size [pc] & 30\,$\times$\,30\,$\times$\,30 & \\
    Resolution [pc] & 0.06 & \\
    Collision velocity [km\,s$^{-1}$] & 5 & \\
    \hline
    Parameter & The small cloud & The large cloud & note\\
    \hline
    Temperature [K] & 120 & 240  \\
    Free-fall time [Myr] & 5.31 & 7.29 \\
    Radius [pc] & 3.5 & 7.2 & \\
    Mass [$M_\odot$] & 417 & 1635 \\
    Velocity dispersion [km\,s$^{-1}$] & 1.25 & 1.71 \\
    Density [cm$^{-3}$] & 47.4 & 25.3 & assumed\\
     & & & a Bonner-Ebert sphere \\
    \hline \hline
  \end{tabular}\label{table:4}
\end{center}
\end{table}

\begin{table}[] 
\begin{center}
  \caption{Observational parameters of the NRO 45\,m telescope dataset \citep{shi2011}}
\def\arraystretch{1.}
\small
   \begin{tabular}{cccccccccccc}
    \hline    \hline
    Diameter & 45\,m& \\
    Date & 2007\,Decl. -- 2008\,May & \\
    Molecular transition & $^{12}$CO($J$\,=\,1--0; 115.271\,GHz) &  \\
    Observation mode & On-the-fly mode &  \\
    Observation time & 40\,hr &  & \\
    Mapping area & 1$\fdg$2\,$\times$\,1$\fdg$2 &  & \\
    FWHM beam size & 15$''$ &  & \\
    Spatial grid size & 7.$^{''}$5 &  & \\
    Effective beam size & 21$''$ &  & \\
    Receiver & BEARS \citep{sun2000, yam2000}\\
    T$_{\rm sys}$ & 250\,--\,500 K in the single-side band &  & \\
    T$_{\rm rms}$ & 0.94\,K &  & \\
    Velocity resolution inside the cloud & 0.5\,km\,s$^{-1}$ &  & \\
    Band width & 32\,MHz &  & \\
    \hline \hline
  \end{tabular}\label{table:3}
\end{center}
\end{table}

\begin{table}[]
\begin{center}
  \caption{The model parameters of the two clouds in M42 and M43}
   \begin{tabular}{cccccccccccc}
    \hline\hline
    Parameter & The blue-shifted cloud & The red-shifted cloud \\
    \hline
    Velocity range [km\,s$^{-1}$] & 4.0\,--\,11.1 & 11.1\,--\,14.9 \\
    Length [pc] & 8.9 & 4.7 & \\
    Width [pc] & 6.5 & 3.4 & \\
    Mass [$M_\odot$] & 1.5\,$\times$\,10$^{4}$ & 3.4\,$\times$\,10$^{3}$ \\
    \hline \hline
  \end{tabular} \label{table:5}
\end{center}
\end{table}

\clearpage


\setcounter{section}{1}

 \begin{figure}
  \epsscale{1.0}
   \plotone{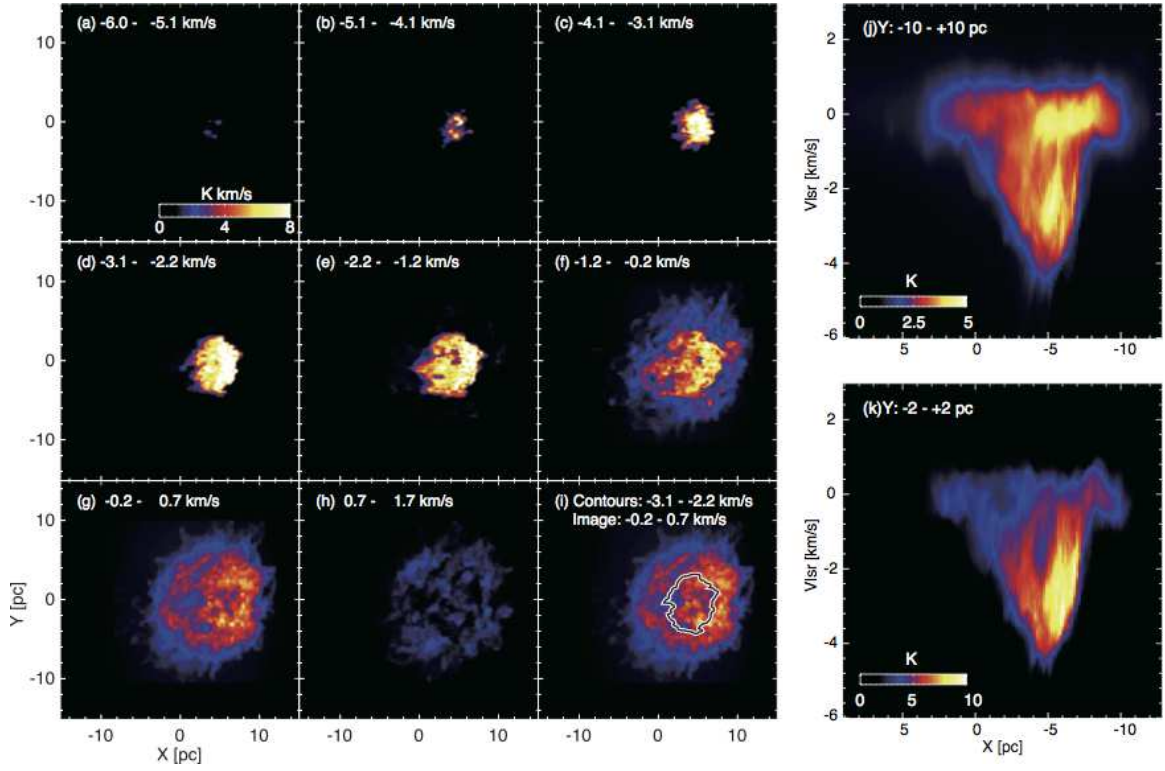}
  \caption{Same as Figure\,\ref{fig3}, but for $\theta=45^\circ$ (Figure\,\ref{fig:takahira}(c)). 
  The image in (i) is the same as (g), and the contours in (d) are plotted at 3\,K\,km\,s$^{-1}$.
   \label{fig4}}
\end{figure}

 \begin{figure}
  \epsscale{0.6}
   \plotone{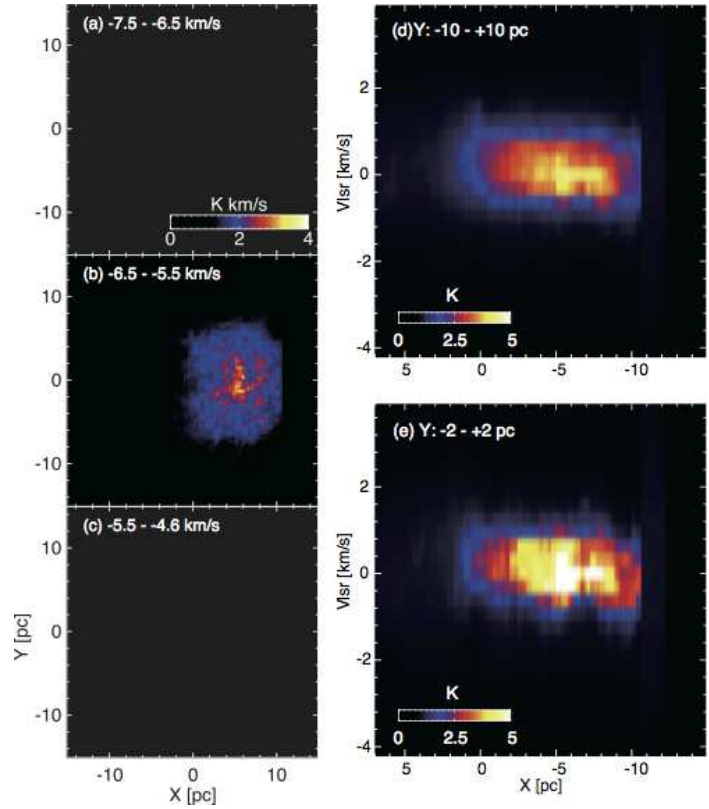}
  \caption{Same as Figure\,\ref{fig3}, but for $\theta=90^\circ$ (Figure\,\ref{fig:takahira}(c)). 
 \label{fig5}}
\end{figure}

 \begin{figure}
  \epsscale{1}
   \plotone{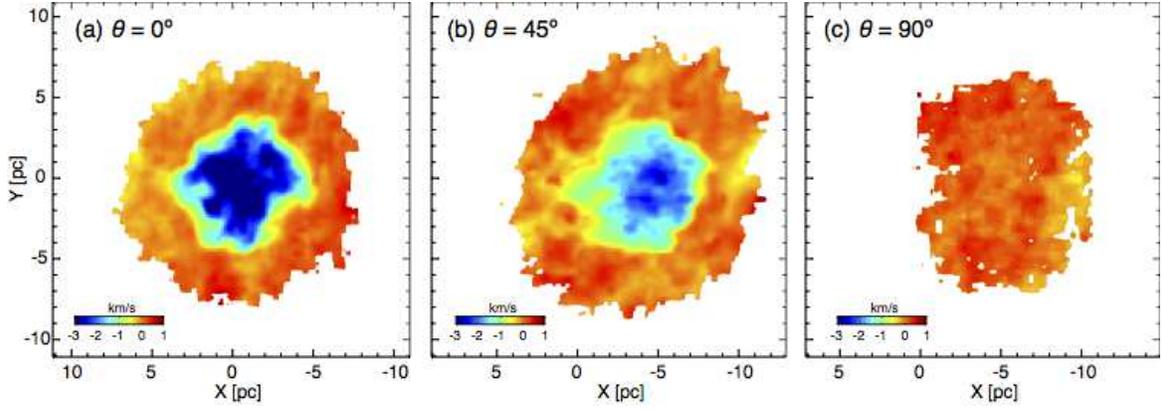}
  \caption{{The first moment maps of the CCC model data for the three inclination angles (a) $\theta$ = $0^\circ$, (b) $45^\circ$, and (c) $90^\circ$.}}
 \label{figmom1}
\end{figure}

 \begin{figure}
  \epsscale{0.65}
   \plotone{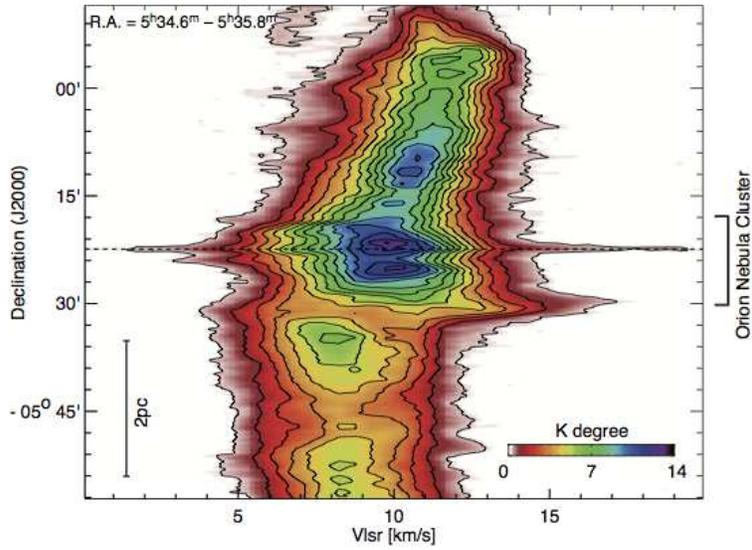}
  \caption{Decl.-velocity diagram of the $^{12}$CO($J$\,=\,1--0) emission toward M42 and M43.
  Contours are plotted every 1\,K\,degree from 0.25\,K\,degree.
  The horizontal dashed line indicates of the position of the Becklin-Neugebauer object at Decl.\,=\,$-5\degr\,22\fm4$.
  \label{fig6}}
\end{figure}

 \begin{figure}
  \epsscale{1.}
   \plotone{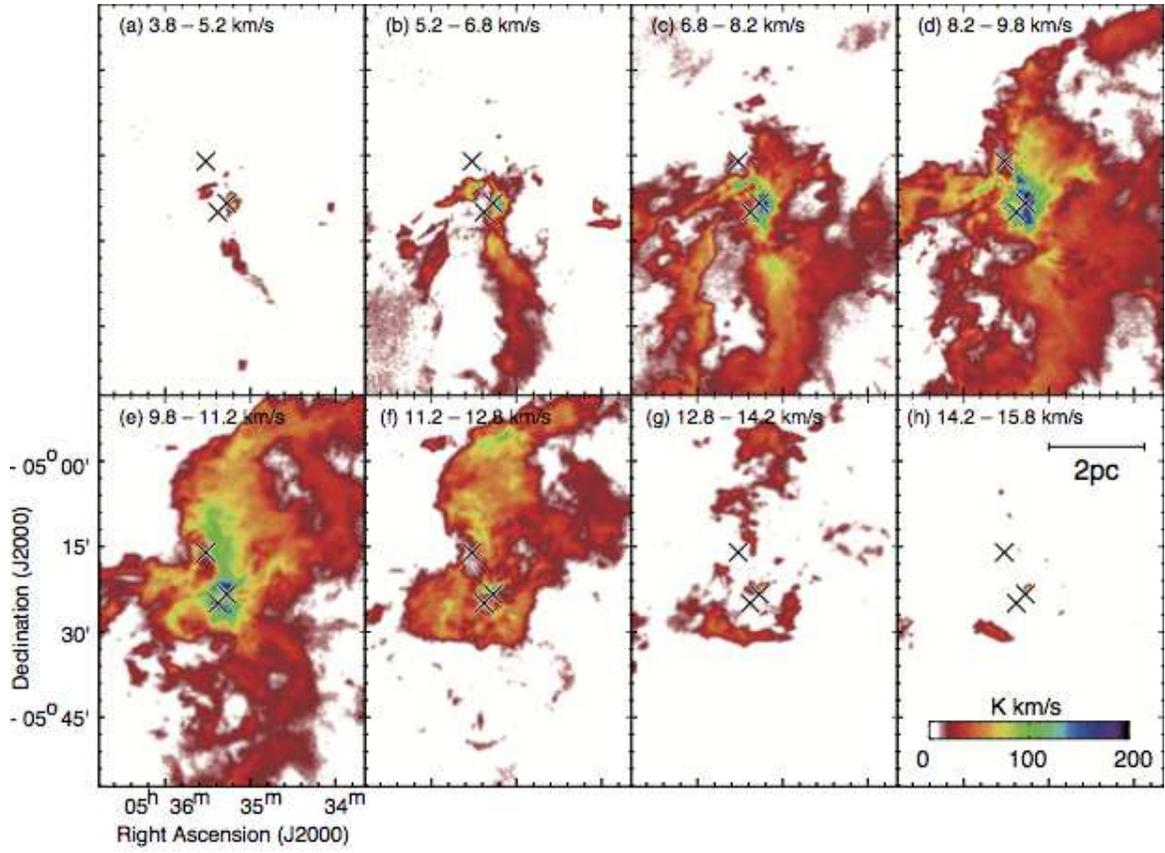}
  \caption{Velocity-channel distributions of the $^{12}$CO($J$\,=\,1--0) emission toward M42 and M43. 
  The false color image indicates the $^{12}$CO($J$\,=\,1--0) integrated intensity. 
  The crosses depict the positions of $\theta^{1}$\,Ori C (right one of the lower right), $\theta^{2}$\,Ori A (left one of the lower right), and NU\,Ori (upper left). 
  \label{fig7}}
\end{figure}

 \begin{figure}
  \epsscale{0.6}
   \plotone{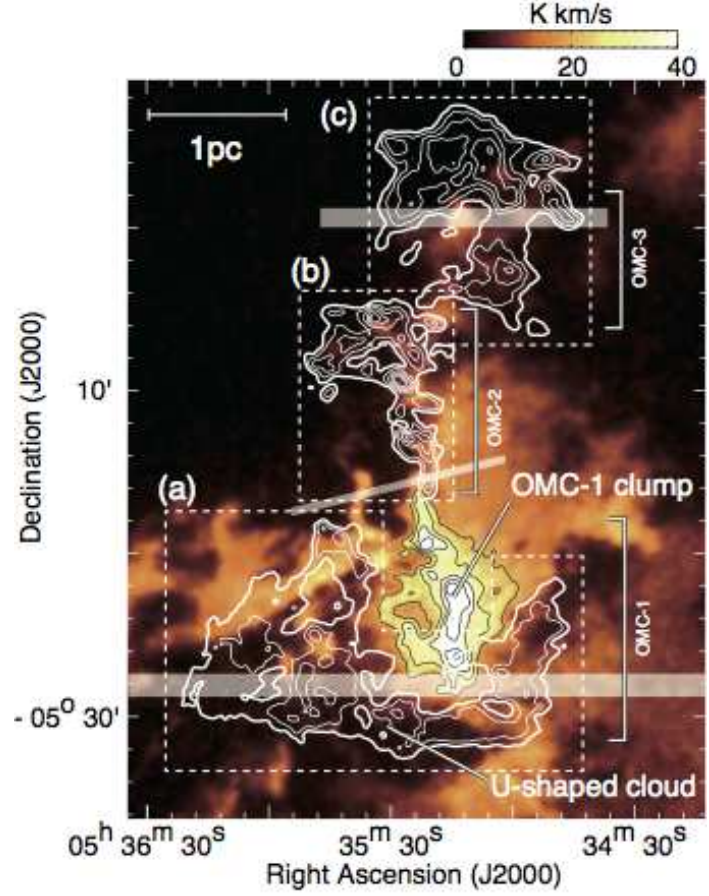}
  \caption{{A comparison between gas distributions of the blue-shifted cloud (image and black contours) and the red-shifted cloud (white contours). The three regions of the complementary distribution are indicates as the three boxes with dashed lines labeled (a), (b) and (c). The velocity range for the blue-shifted cloud is 8.8 km s$^{-1}$, while those for the red-shifted cloud are 12.9 km s$^{-1}$, 12.9 -- 14.9 km s$^{-1}$, and 12.9--14.9 km s$^{-1}$ in (a), (b), and (c), respectively. The lowest level and internal of the black contours for the blue-shifted cloud are 21 K km s$^{-1}$ and 7 K km s$^{-1}$, respectively, while those of the white contours for the red-shifted cloud are plotted at (a) 5.5 K km s$^{-1}$ and 4 K km s$^{-1}$, (b) 13 K km s$^{-1}$ and 7 K km s$^{-1}$, and (c) 14 K km s$^{-1}$ and 8 K km s$^{-1}$.}
 \label{fig8}}
\end{figure}

 \begin{figure}
  \epsscale{0.65}
   \plotone{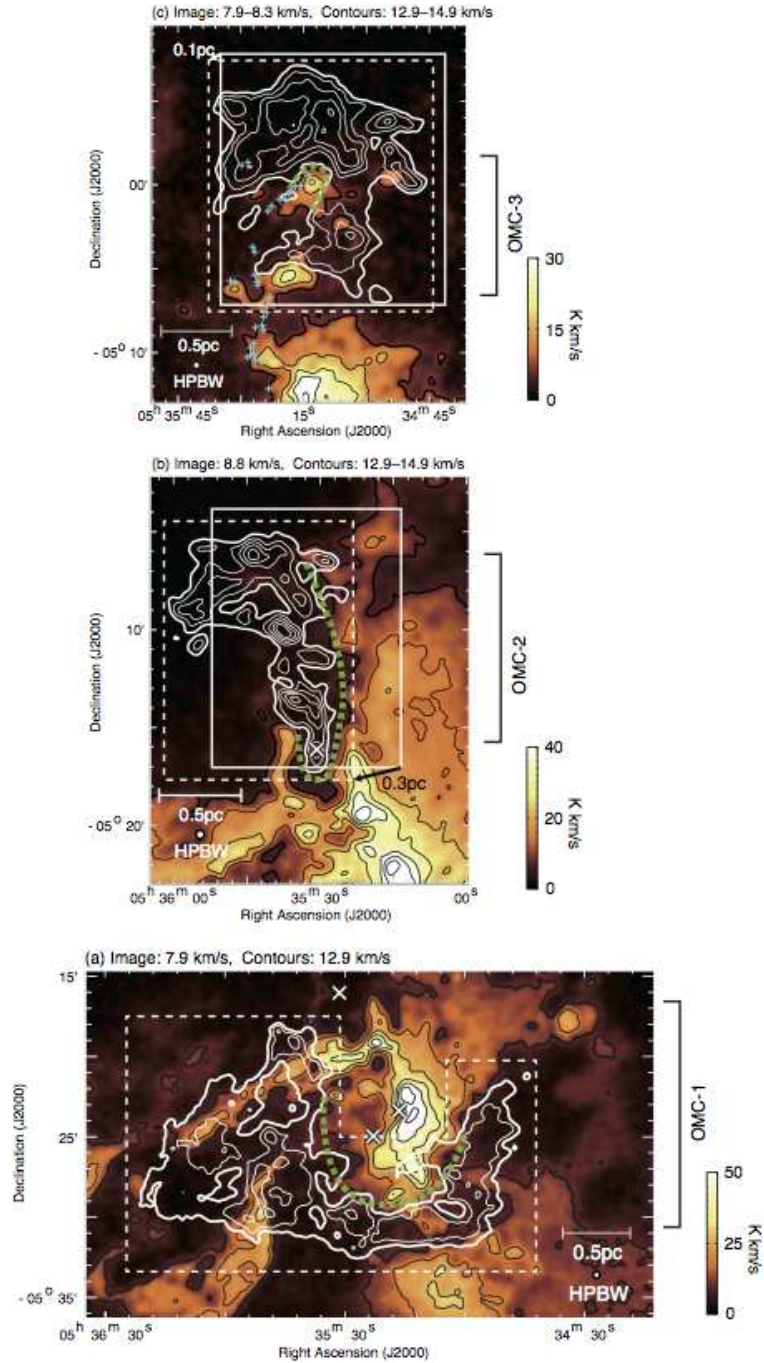}
  \caption{The complementary distributions of the two velocity distributions are shown in Fig. 10(a), (b), and (c) from the bottom to the top after displacement shown by arrows for (b) and (c) with small tuning in velocity ranges. The image with black contours and the white contours indicate the blue- and red-shifted clouds, respectively. {The contacting faces of the two clouds in the complementary distributions are indicated as thick dashed lines colored in green.} The velocity ranges for the blue-shifted cloud and the red-shifted cloud; (a) 7.9\,km\,s$^{-1}$ and 12.9\,km\,s$^{-1}$, (b) 8.8\,km\,s$^{-1}$ and 12.9\,--\,14.9\,km\,s$^{-1}$, and (c) 7.9\,--\,8.3\,km\,s$^{-1}$ and 12.9--14.9\,km\,s$^{-1}$. {The lowest level and internal of the white contours are (a) 5.5\,K\,km\,s$^{-1}$ and 4\,K\,km\,s$^{-1}$, (b) 13\,K\,km\,s$^{-1}$ and 7\,K\,km\,s$^{-1}$, and (c) 14\,K\,km\,s$^{-1}$ and 8\,K\,km\,s$^{-1}$, while those of the black contours are (a) 8\,K\,km\,s$^{-1}$ and 7\,K\,km\,s$^{-1}$, (b) 7\,K\,km\,s$^{-1}$ and 7\,K\,km\,s$^{-1}$, and (c) 7\,K\,km\,s$^{-1}$ and 7\,K\,km\,s$^{-1}$.} In (c) blue crosses show low-mass young stars in OMC-2\,/\,3 \citep{pet2008}, while in (a) and (b) NU\,Ori, $\theta^{1}$\,Ori C and $\theta^{2}$\,Ori A are plotted with white crosses.
 \label{fig9}}
\end{figure}

 \begin{figure}
  \epsscale{0.6}
   \plotone{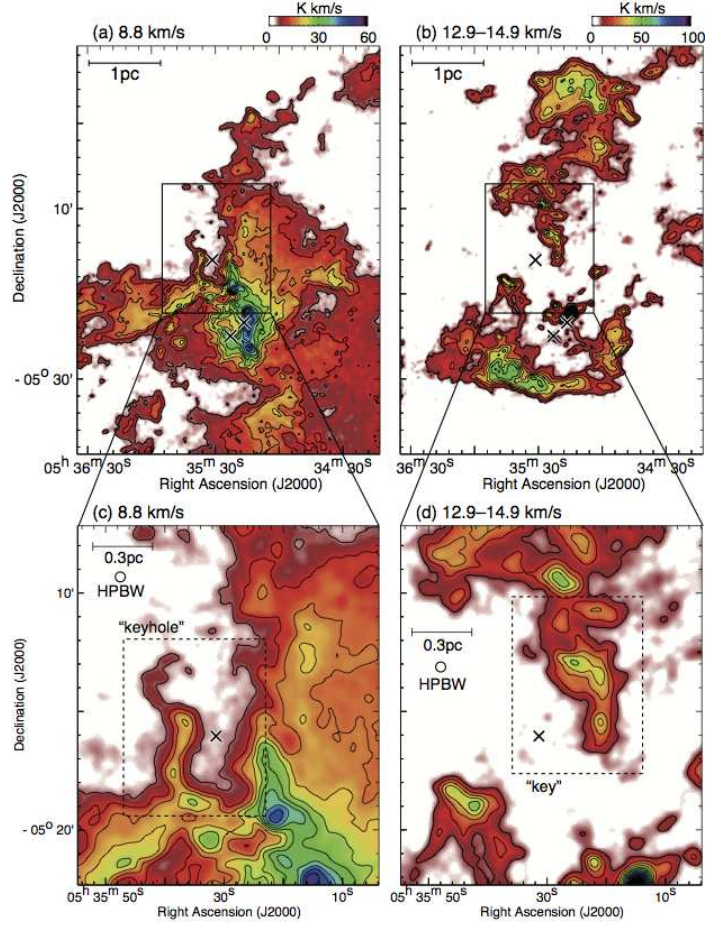}
  \caption{The typical $^{12}$CO($J$\,=\,1--0) distributions of the blue-shifted and the red-shifted clouds toward the ONC, where the velocities of the contours and the image are 8.8\,km\,s$^{-1}$ for (a) and (c), and 12.9\,--\,14.9\,km\,s$^{-1}$ for (b) and (d). 
  The crosses denote $\theta^{1}$\,Ori C, $\theta^{2}$\,Ori A and NU\,Ori as in Figure 7. 
  (c) and (d) indicate the detailed distribution, which include the Orion $``$Keyhole$"$ and $``$Key$"$, respectively, in the box shown in (a) and (b). 
  Contours are plotted every 5\,K\,km\,s$^{-1}$ for the blue-shifted cloud, and every 8\,K\,km\,s$^{-1}$ from 10\,K\,km\,s$^{-1}$ for the red-shifted cloud.
   \label{fig10}}
\end{figure}

 \begin{figure}
  \epsscale{.9}
   \plotone{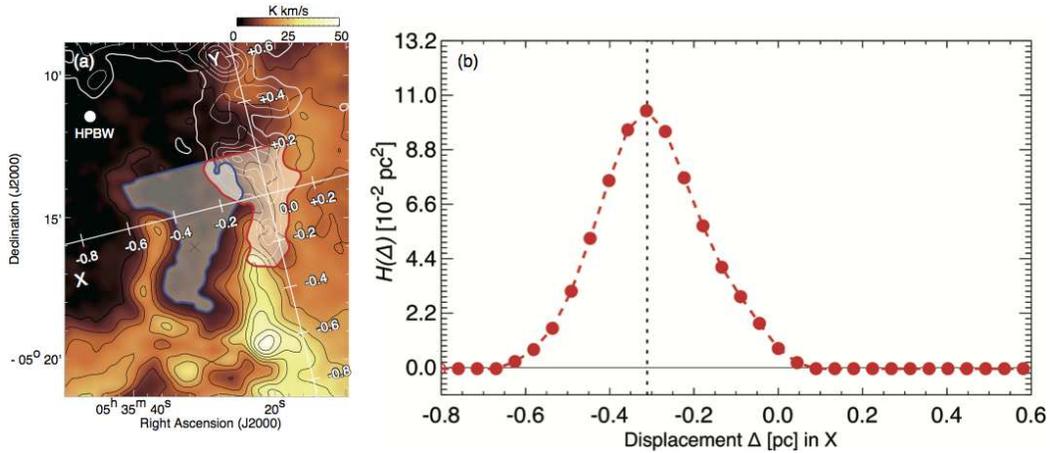}
  \caption{(a) The image and the blue contours shows the blue-shifted cloud including the Orion Keyhole, and the white and red contours show the red-shifted cloud including the Orion Key. The blue and red contours with white shaded region indicate the lowest intensity limit 5 and 9\,K\,km\,s$^{-1}$ of the Keyhole and the Key, respectively. The contour levels are the same as those in Figure\,10. X-axis was taken for the direction of the displacement of the Orion Key as explained in the text and the Y-axis is normal to the X-axis. See the text and the Appendix for more details. 
  (b) the overlapping integral $H(\Delta)$ for the Key and Keyhole in the Appendix is presented, where the Orion Key and Keyhole were assumed to be a uniform value 1.0 (arbitrary unit) with a grid spacing of 0.1\,pc\,$\times$\,0.1\,pc. 
   \label{fig11}}
\end{figure}
\clearpage

 \begin{figure}
  \epsscale{.45}
   \plotone{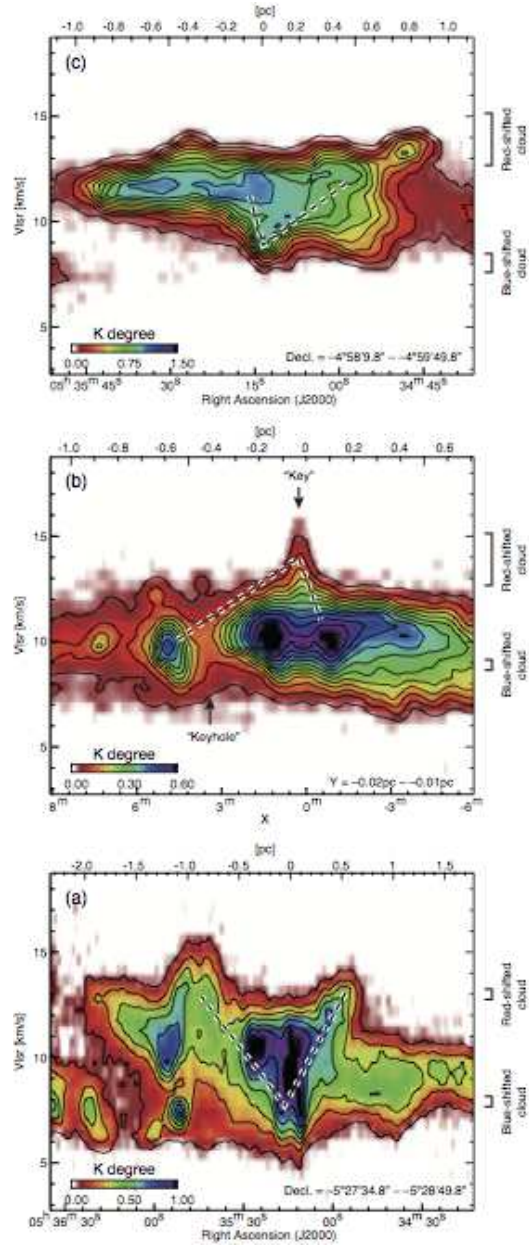}
  \caption{{Position-velocity diagrams of the $^{12}$CO($J$=1--0) emissions in the three regions (a)--(c) in Figure\,\ref{fig8}, where the integration ranges are indicated as filled white areas in Figure\,\ref{fig8}. While in (a) and (c) gas distributions along R.A. are presented, while (b) shows that along the X-axis defined in Figure\,\ref{fig11}(a). The velocity ranges of the blue-shifted and red-shifted clouds in the integrated intensity maps in Figure\,\ref{fig9} are shown on the right sides of the panels. }
   \label{fig:pv_abc}}
\end{figure}
\clearpage

 \begin{figure}
  \epsscale{1}
   \plotone{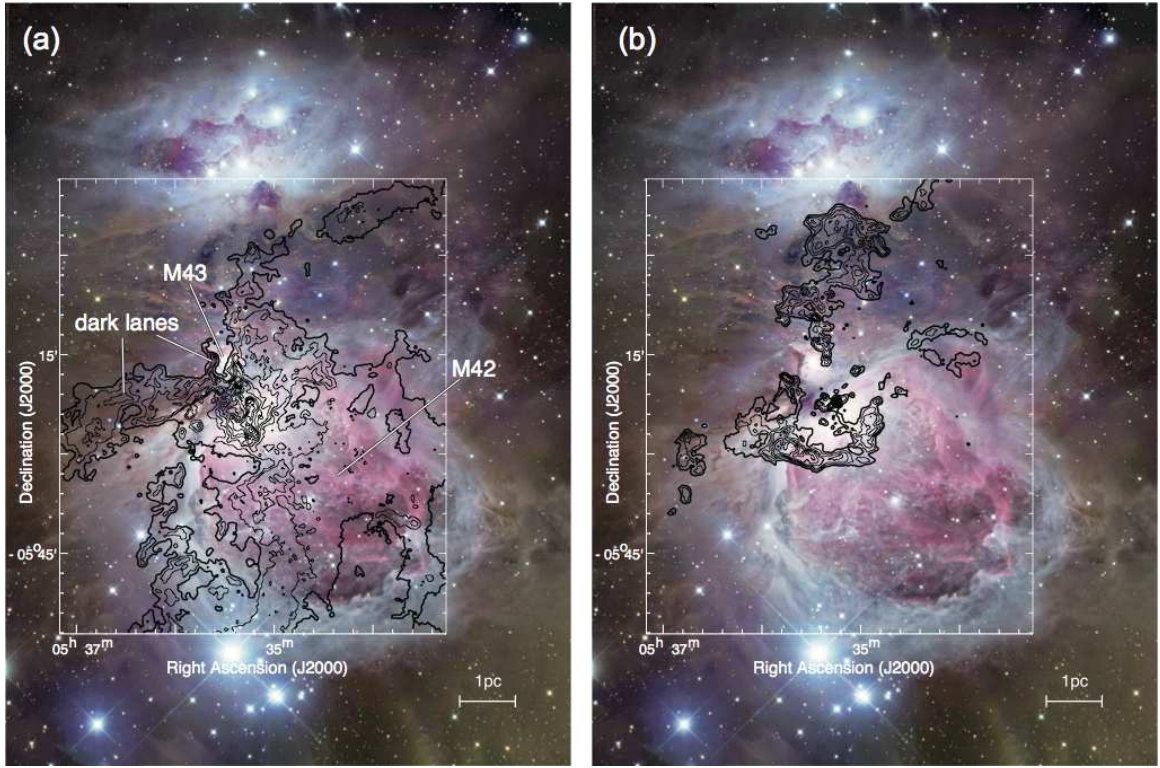}
  \caption{The contour map of (a) the blue-shifted cloud at 8.8\,km\,s$^{-1}$ {and (b) the red-shifted cloud at 12.9\,--\,14.9\,km\,s$^{-1}$} are superimposed on the optical image of M42 and M43. 
 Contours are plotted every 5\,K\,km\,s$^{-1}$. Image courtesy of Robert Gendler.
  \label{fig12}}
\end{figure}

 \begin{figure}
  \epsscale{1}
   \plotone{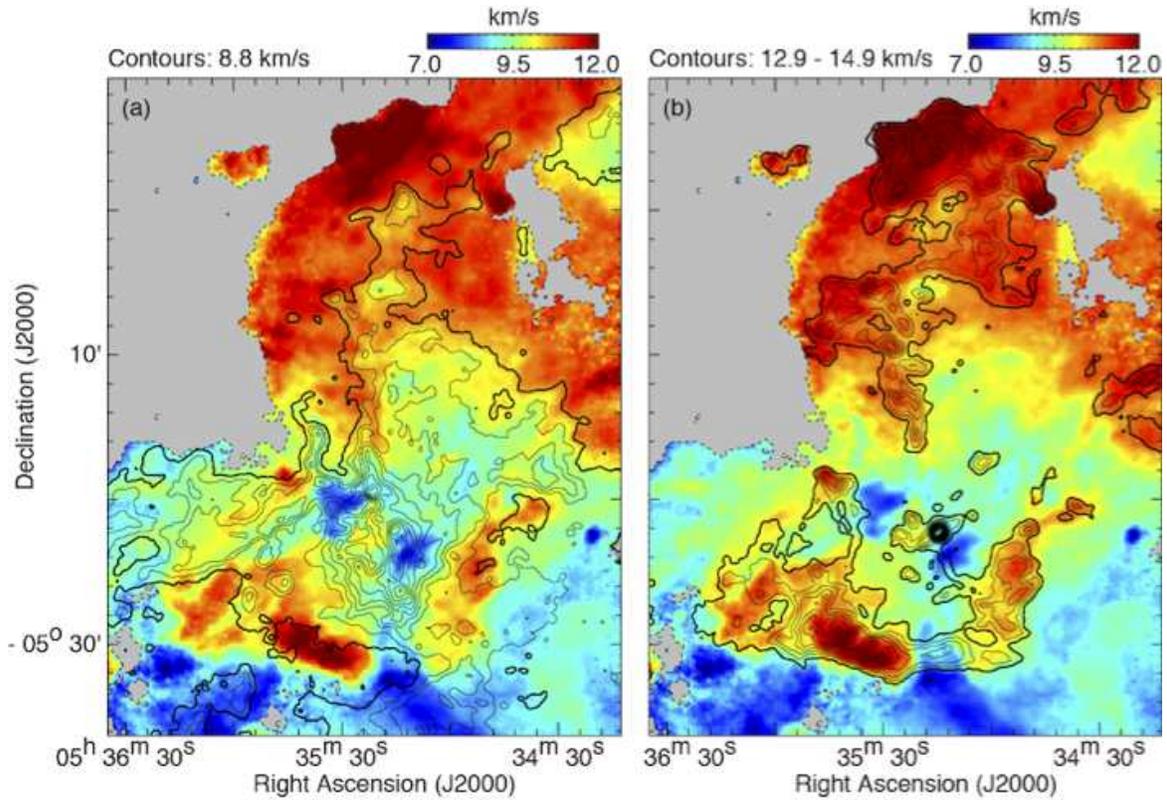}
  \caption{{
The first moment map is presented, where the contour maps in (a) and (b) show the blue- and red-shifted clouds, respectively. The first moment map was made for a velocity range between -7 and 16\,km\,s$^{-1}$ using the velocity channels having CO intensities larger than 2K. Contours are plotted every 5\,K\,km\,s$^{-1}$ for the blue-shifted cloud, and every 8\,K\,km\,s$^{-1}$ from 10\,K\,km\,s$^{-1}$ for the red-shifted cloud.}
\label{fig-ori-mom1}}
\end{figure}

 \begin{figure}
  \epsscale{0.6}
   \plotone{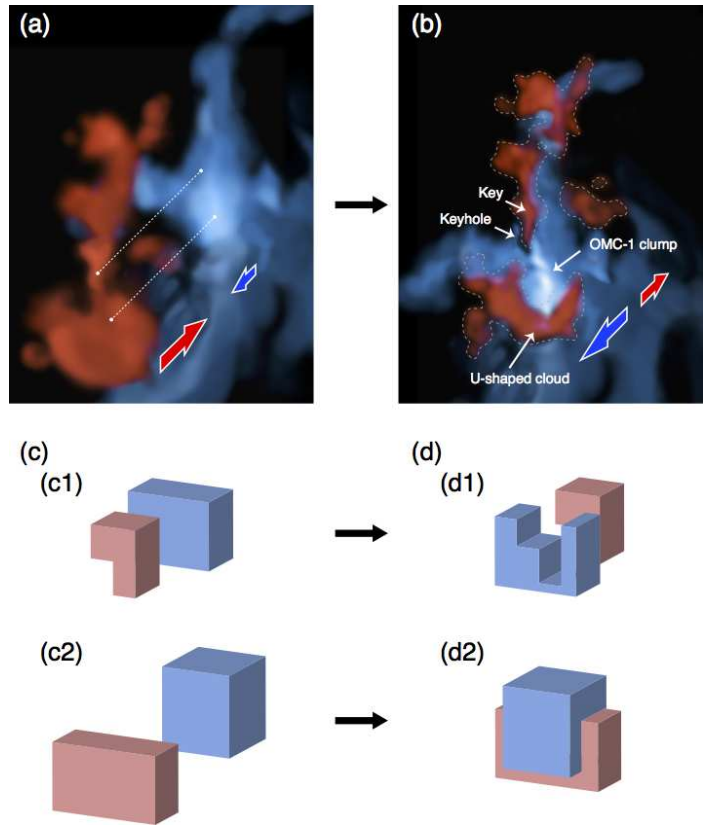}
  \caption{
  The upper left image (a) shows the two clouds before the collision and the upper right image (b) after the collision projected in the sky. 
  The lower left panels show rectangular-solid model clouds before the collision (c1) for the Key and Keyhole and (c2) for the M42 clump and the U-shape cloud, and the lower right panels after the collision (d1) for the Key and Keyhole and (d2) for the M42 clump and the U-shape cloud.
    \label{fig14}}
\end{figure}

 \begin{figure}
  \epsscale{.9}
   \plotone{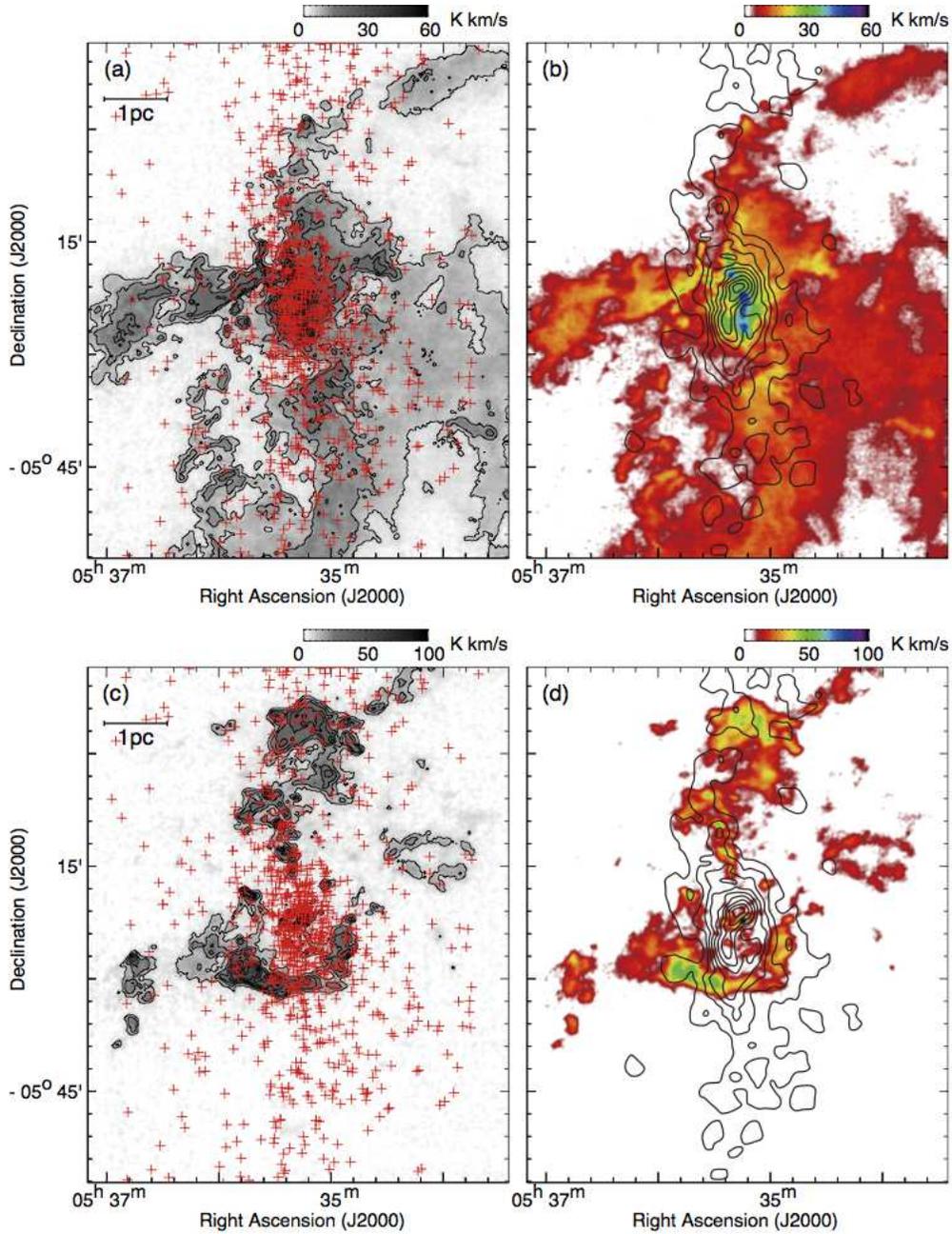}
  \caption{Comparisons between the $^{12}$CO($J$\,=\,1--0) distribution of the blue-shifted cloud and 2MASS variable stars \citep{car2001}.
   (a) The image and contours indicate the blue-shifted cloud at 8.8\,km\,s$^{-1}$, and the red crosses indicate variable stars. 
   (b) The image indicates the blue-shifted cloud, and the contours indicate density distributions of the 2MASS stars.
   {(c) and (d) are same as (a) and (b), but for the red-shifted cloud at 12.9--14.9\,km\,s$^{-1}$.}
   The contour levels are from 2.3\,$\times$\,10$^{2}$\,stars\,pc$^{-2}$ to 2.8\,$\times$\,10$^{3}$\,stars\,pc$^{-2}$ every 2.9\,$\times$\,10$^{2}$\,stars\,pc$^{-2}$ 
  \label{fig15}}
\end{figure}

\clearpage
\appendix
\section{Estimate of the displacement in the complementary distribution}
 
\begin{figure}
  \epsscale{0.6}
   \plotone{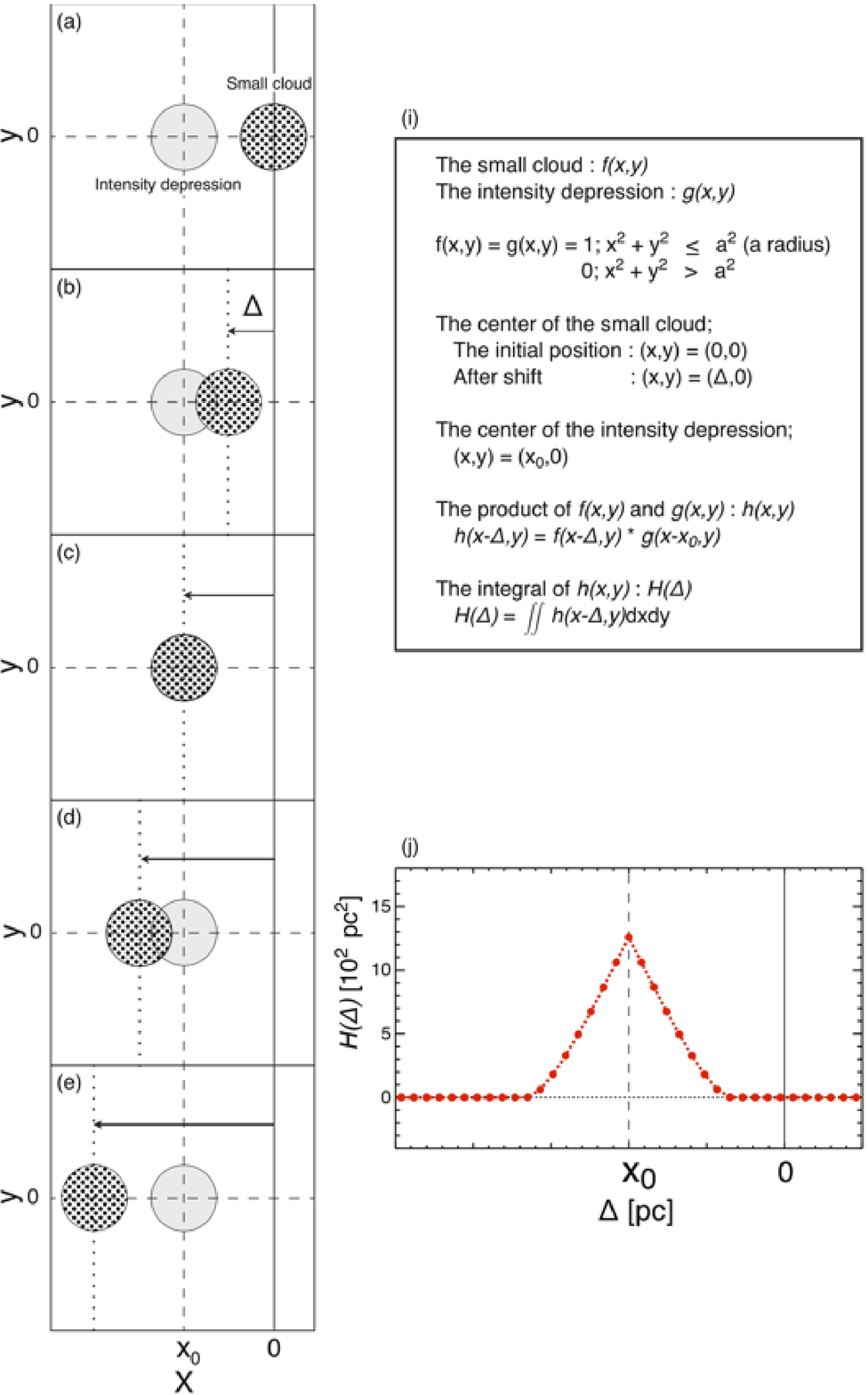}
  \caption{A simple model of the small cloud and the intensity depression. 
  The both are assumed to be a circle of radius a, where we take a\,=\,1\,pc. 
  In order to calculate the overlapping integral $H(\Delta)$, the shift of the small cloud $\Delta$ is swept along the X-axis as shown in the left five panels (a)--(e), where (a) is the initial state. 
  In the upper right (i), the equations are summarized, and in the lower right (j) $H(\Delta)$ is shown. 
  The peak of $H(\Delta)$ gives the initial displacement $x_{0}$ of the two circles.
  \label{figa2}}
\end{figure}

We describe the method for a simple case of the complementary distribution in the X-Y plane (Figure\,\ref{figa2}) where the small cloud and the intensity depression are assumed to be circles of the same radius a with an initial displacement of $x_{0}$ along the X-axis in the sky. By shifting the small cloud by $\Delta$ in the direction of the intensity depression along the X-axis, we calculate the overlapping product between the two circles represented as $f(x,y)$ and $g(x,y)$ by assuming that they have uniform intensity 1.0 in an arbitrary unit inside the circle and 0.0 outside the circle. By taking a product of the circle and the intensity depression at $\Delta$, the overlapping between the two circles is given by $h(x-\Delta,y)$, the product of $f(x-\Delta,y)$ and $g(x-x_{0},y)$. $h(x-\Delta)$ is integrated in the X-Y plane and is expressed by $H(\Delta)$. The displacement of the two circles is then estimated to be $\Delta\,=\,X_{0}$ at the peak of $H(\Delta)$ (Figure\,\ref{figa2}).

The method is applied to the synthetic observations in Figures\,\ref{figa2}(b), (c), and (d).  In Figures\,\ref{figa2}(b) and (c), the distributions of the small cloud and the intensity depression are taken from Figure\,\ref{fig4}. Grid spacing is given as 0.1\,pc\,$\times$\,0.1\,pc in the X-Y plane according to the numerical simulations \citep{tak2014}, and we assign a numerical value to be 1.0, for each pixel inside the both components, and to be zero outside of them as above (Figures\,\ref{figa2}(i) and (j)). $H(\Delta)$ is calculated in the same manner as above with an incremental step of 0.1\,pc in $\Delta$ and is shown in Figure\,\ref{figa2}(d). The peak position of $H(\Delta)$ gives the displacement to be $\sim$\,2\,pc. By correcting for the projection effect, the displacement corresponds to the length of the cavity $\sim$\,3\,pc, where the size of the small cloud is taken into account. We find the value is roughly consistent with the length of the cavity in the large cloud, $\sim$\,4\,pc, at 1.6\,Myr in the numerical simulations. In Section 3, we applied the method to estimate the displacement between the Orion Key and Keyhole.

\end{document}